\documentclass[twocolumn]{aastex7}

\newcommand{\dqmix}{mixed dipole--quadrupole }

\usepackage{amsmath}

\begin{document}
\raggedbottom

\title{First-Principles Polar-Cap Currents in Multipolar Pulsar Magnetospheres}

\author[0000-0001-6406-1003]{Chun Huang}\email{chun.h@wustl.edu}
\affiliation{Physics Department and McDonnell Center for the Space Sciences, Washington University in St. Louis; MO, 63130, USA;\\}
\correspondingauthor{Chun Huang}
\begin{abstract}
X-ray pulse-profile modeling of millisecond pulsars offers a direct route to measuring neutron star masses and radii, thereby constraining the dense-matter equation of state. However, standard analyses typically rely on \emph{ad hoc} hotspot parameterizations rather than self-consistent physical models. While connecting surface heating directly to the magnetospheric geometry provides a more natural physical pathway, computing global magnetospheric solutions is too computationally expensive to perform on-the-fly during parameter inference. In this work, we bridge this gap by deriving fully analytic, first-principles expressions for surface return currents in mixed dipole--quadrupole magnetospheres. Working within force-free electrodynamics, we generalize the field-aligned current invariant $\Lambda$, the crucial scalar that maps the far-zone magnetic structure to the near-zone heating rate, from the standard dipole approximation to arbitrary quadrupolar configurations. We demonstrate that even when the quadrupole component is sub-dominant in the far zone (the mixing regime), using a dipole-based heating prescription fails to capture the significant enhancement or suppression of the return-current density on the polar cap. Our consistent quadrupole-aware framework reveals that these multipolar currents redistribute the surface heating, leading to systematic discrepancies in predicted pulse profiles that are amplified by atmosphere beaming and can reach $\sim 30\%$ near pulse peaks. These results provide a rigorous analytic foundation for mapping global magnetic geometry to surface heating in multipolar magnetospheres, enabling physically consistent inference beyond the idealized dipole approximation.
\end{abstract}

\section{Introduction}
\label{introduction}

Rotation-powered millisecond pulsars (MSPs) are prime targets for X-ray pulse-profile modeling \citep{Bogdanov_19_dataset,Guillot_19_dataset}. In this framework, thermal emission from surface hotspots is used to infer neutron-star masses, radii, and surface hotspot configuration, thereby providing important constraints on the dense-matter equation of state and the physics of neutron-star interiors \citep{Pechenick_1987,Beloborodov_2002,Poutanen_2006,Morsink_2007,Cadeau_2007,Psaltis_2014,Watts_2016,Nattila_2018,Poutanen_2020}. Thanks to NASA's Neutron Star Interior Composition Explorer (NICER), over the past half-decade such analyses have begun to probe both the dense-matter equation of state and the physics of surface emission in a quantitative way \citep{Bogdanov_19,Bogdanov_21,Miller2019,Riley2019,Miller2021,Riley2021,Salmi_2024_j0740,Choudhury24,Vinciguerra24,Dittmann24,Miller-J0437}.

Existing modeling of pulsars such as PSR~J0030+0451 and PSR~J0437--4715 strongly suggests that the external magnetic-field configurations of these MSPs can deviate significantly from a pure inclined dipole and may include substantial multipolar components \citep{Miller2019,Riley2019,Bilous_2019,Kalapotharakos_2021,Chen_2020,Choudhury24,Petri:2025nhu}. For example, the hotspot pattern inferred for PSR~J0030+0451 indicates a highly non-antipodal configuration that is naturally explained by a mixed dipole--quadrupole field or an off-centered dipole \citep{Kalapotharakos_2021,Chen_2020,Huang:2025_hotspot}. Despite these indications, most current pulse-profile modeling studies still rely on \emph{ad hoc} phenomenological hotspot configurations that are not derived from the underlying magnetic geometry and therefore lack a self-consistent connection between the global magnetic field, polar-cap currents, and the resulting hotspot structure. It is therefore natural to pursue \emph{physics-motivated} hotspot models that begin from an assumed magnetic geometry and then compute the associated return currents and surface heating, rather than fitting arbitrary hotspot configurations. A first step in this direction has been taken with shifted-dipole models \citep{Huang:2025_hotspot} that extend the semi-analytic prescription for inclined dipoles developed by \citet{Gralla2017,Lockhart2019}. However, once higher multipole components are allowed, any serious attempt at physics-motivated pulse-profile modeling must systematically relate a multipolar magnetic field to the surface current distribution and the resulting hotspot temperature map. Such a framework must be computationally efficient to enable on-the-fly inference while ensuring self-consistency between the far-zone boundary conditions and the near-zone heating.

A powerful semi-analytic framework for connecting global magnetic geometry to polar-cap currents and surface temperature distributions was developed by \citet{Gralla2017,Lockhart2019}. This approach utilizes a spacetime formulation of force-free electrodynamics and matched asymptotic expansions, organized in powers of the compactness parameter $\epsilon$. A central outcome is the identification of a conserved scalar $\Lambda(\alpha,\beta)$ on magnetic field sheets, which governs the field-aligned current near the star \citep{Mestel_1973,Beskin_1983,Uchida_1998,Gruzinov_2005}. Previously, the explicit form of $\Lambda$ was obtained by fitting time-dependent 3D simulations of oblique dipoles \citep{Gralla2017}. While useful, this relies on a \emph{dipole-only} matching condition, implicitly assuming that the magnetic field is purely dipolar in the overlap region where the near and far zones meet.

A common justification for this restriction is that higher multipoles decay rapidly in the outer magnetosphere. As an order-of-magnitude estimate, a vacuum multipole of degree $\ell$ decays as $B_\ell \propto r^{-(\ell+2)}$, so the quadrupole-to-dipole ratio scales as
\begin{equation}
\frac{B_Q}{B_D}(R_{\rm LC}) \sim
\left.\frac{B_Q}{B_D}\right|_{R_\star}
\left(\frac{R_\star}{R_{\rm LC}}\right).
\end{equation}
For MSPs, where $\epsilon \equiv R_\star/R_{\rm LC} \sim 10^{-2}$--$10^{-1}$, this implies that even if the surface field has comparable dipole and quadrupole components, the quadrupole contribution in the far zone is suppressed. Indeed, this is compatible with gamma-ray light curves that are often well reproduced by effectively dipolar geometric models \citep{Li_2011,Bai_2010,harding_2016,Watts_2016,Romani_2010,Johnson_2014}.

However, this far-zone dominance does not justify ignoring multipoles in the matching region. The scaling above indicates that a substantial quadrupole component persists in the overlap region at the level of a few to tens of percent. Crucially, this is precisely the region where the invariant $\Lambda$ is fixed. Consequently, using a far-zone construction that enforces a strict dipole introduces a systematic mismatch in the predicted {magnetospheric current density evaluated at the stellar surface}. This initial error is physically significant because it modifies the return-current density, which is then further processed by atmospheric radiative transfer. Realistic atmospheric beaming can amplify moderate current discrepancies into large deviations ($\sim 30\%$) in the observed X-ray waveform. A physically consistent heating model therefore requires matching that admits multipolar structure in the fixing region and derives $\Lambda$ from first principles.

In this work we address this theoretical gap. Working within the same force-free framework and at the same order in $\epsilon$ as \citet{Gralla_2016,Gralla2017,Lockhart2019}, we generalize the leading-order magnetic geometry from a pure dipole to a dipole--quadrupole configuration. We derive fully analytic expressions for the conserved scalar $\Lambda(\alpha,\beta)$ along each field sheet, without relying on numerical fitting. We first obtain a first-principles analytic expression for the inclined dipole and then extend the derivation to the axisymmetric $m=0$ quadrupole and, more generally, to arbitrary quadrupolar components. Once the multipolar geometry and associated Euler potentials are specified, the far-zone matching condition uniquely fixes the field-aligned current invariant $\Lambda$. Consequently, our results yield explicit expressions that define the open-field region, set the polar-cap current density, and govern the resulting surface temperature map in mixed dipole--quadrupole geometries. This connects multipolar magnetic geometry to polar-cap currents and hotspot heating, allowing pulse-profile modeling to incorporate dipole–quadrupole structure without requiring a global force-free simulation at each parameter choice.

The remainder of this work is organized as follows. In Section~\ref{sec:Lambda-derivation} we state our assumptions, review the force-free setup, and formulate the problem in terms of Euler potentials and the conserved scalar $\Lambda(\alpha,\beta)$. We then present first-principles, analytic-form derivations of $\Lambda$ for the inclined dipole, for the axisymmetric ($m=0$) quadrupole, and for a general quadrupolar field, including mixed dipole--quadrupole configurations. In Section~\ref{result} we present the resulting polar-cap current distributions and surface temperature profiles, and use them to compute representative X-ray pulse profiles, quantifying the physical impact of multipolar corrections on the observable signal, especially considering realistic atmospheric beaming. In Section~\ref{sec:discussion} we discuss the limitations of our framework. In Section~\ref{conclusion} we summarize our findings and outline directions for future work.
\section{Method and formalism}
\label{sec:Lambda-derivation}

As shown by \citet{Gralla2017}, stationary, ideal force-free fields admit a conserved scalar $\Lambda(\alpha,\beta)$ on magnetic field sheets. This scalar fixes the component of the four-current parallel to the magnetic field and therefore controls the return-current distribution on the polar cap. Throughout we refer to $\Lambda$ as the \emph{field-aligned current invariant} and define it by
\begin{equation}
\Lambda \equiv \frac{\mathbf{J}\cdot\mathbf{B}}{\mathbf{B}^2},
\label{eq:Lambda_def}
\end{equation}
so that the parallel-current contribution is $\mathbf{J}_\parallel=\Lambda\,\mathbf{B}$. While the full current density generally includes a perpendicular component, we focus here on the field-aligned contribution controlled by $\Lambda$, as $\mathbf{J}_\parallel$ governs the return currents responsible for heating the stellar surface.

In general, obtaining analytic-form expression for $\Lambda(\alpha,\beta)$ on the stellar polar cap is non-trivial. Nevertheless, such an expression enables one to connect an assumed magnetic geometry to hotspot heating and X-ray pulse profiles without running a global force-free simulation for each parameter choice. In the framework of \citet{Gralla2017} and its application to dipolar and axisymmetric dipole--quadrupole geometries by \citet{Lockhart2019}, the existence of $\Lambda$ follows whenever the electromagnetic field is stationary and ideal, i.e.\ $F\wedge F=0$ and $\xi\cdot F=0$ for the Killing field $\xi$. However, the explicit cap prescription used previously was calibrated by fitting to time-dependent force-free simulations of a dipolar far-zone magnetosphere, rather than derived analytically, and a systematic extension to more general multipolar fields was not provided.

In this section we address this problem within the same matched-asymptotic, force-free framework as \citet{Gralla_2016,Gralla2017,Lockhart2019}, but allowing for a more general leading-order magnetic geometry. We first derive a fully analytic expression for $\Lambda(\alpha,\beta)$ in the inclined-dipole case, and explain how it relates to the semi-analytic prescription adopted by \citet{Gralla2017}. We then extend the derivation to an axisymmetric ($m=0$) quadrupole and, more generally, to an arbitrary quadrupolar contribution, thereby obtaining $\Lambda$ for mixed dipole--quadrupole fields in closed form. We close the section by summarizing how these ingredients are used to construct hotspot temperature distributions and to compute X-ray light curves for pulse-profile modeling.

\subsection{Problem setup}

Within force-free electrodynamics, we consider a stationary, ideal force-free pulsar magnetosphere. The force-free condition implies that the electromagnetic field is degenerate, so that $F\wedge F=0$. In this case there exist Euler potentials $\alpha$ and $\beta$ such that the field can be written as
\begin{equation}
F = d\alpha \wedge d\beta .
\end{equation}
To leading order in the small parameter $\epsilon$ that controls the matched-asymptotic expansion, the field-aligned current density can be expressed in terms of a scalar function $\Lambda(\alpha,\beta)$ that is constant on each magnetic field sheet. This scalar provides an effective description of the current distribution and is guaranteed to exist under the conditions stated in \citet{Gralla2017} for stationary, ideal force-free field.

The open-flux polar-cap region is assumed to be a small, simply connected domain on the stellar surface. It is therefore convenient to introduce local polar coordinates $(\rho,\varphi')$ on each cap. At the order considered here, curvature and general-relativistic factors vary only weakly across a cap and can be treated as approximately constant. Inside the cap, $\Lambda$ is taken to be smooth and finite up to the boundary, while the thin return-current layer along the separatrix is treated separately. Global current closure between the two hemispheres is imposed explicitly.

\subsection{Oblique dipole case}
\label{subsec:oblique-dipole}

With the electromagnetic field tensor written in terms of Euler potentials as $F=d\alpha\wedge d\beta$, the force-free Ohm's law takes the form
\begin{equation}
\rho_e \mathbf{E}+\mathbf{J} \times \mathbf{B}=0, 
\qquad 
\mathbf{E} \cdot \mathbf{B}=0 .
\end{equation}
Here $\rho_e$ is the charge density, $\mathbf{E}$ is the electric field, $\mathbf{B}$ is the magnetic field, and $\mathbf{J}$ is the electric current density (measured by corotating observers). The first relation expresses the vanishing of the Lorentz-force density, while $\mathbf{E}\cdot\mathbf{B}=0$ states that the electric field is everywhere orthogonal to the magnetic field in the force-free region.

Solving the force-free Ohm's law algebraically for $\mathbf{J}$ gives
\begin{equation}
\mathbf{J}
=
\frac{\rho_e \mathbf{E} \times \mathbf{B}}{B^2}
+
\frac{\mathbf{J} \cdot \mathbf{B}}{B^2} \mathbf{B},
\label{eq:solution_J}
\end{equation}
where $B^2 \equiv \mathbf{B}\cdot\mathbf{B}$. The first term describes the drift current perpendicular to $\mathbf{B}$, and the second term is the field-aligned component of $\mathbf{J}$.

{Using the invariant $\Lambda \equiv(\mathbf{J}\cdot\mathbf{B})/\mathbf{B}^2$ defined in equation \eqref{eq:Lambda_def}, equation \eqref{eq:solution_J} may be written as
\begin{equation}
\mathbf{J}=\Lambda \mathbf{B}+\rho_e \frac{\mathbf{E} \times \mathbf{B}}{B^2} .
\label{eq:JinLambda}
\end{equation}
The first term is purely field-aligned, while the second represents the perpendicular drift current.

{Throughout this work, the three-current $\mathbf{J}$ and charge density $\rho_e$ are those measured by corotating observers on the stellar surface and in the near zone. In \citet{Gralla2017}, the invariant $\Lambda$ is equivalently introduced via the decomposition of the current in the fixed frame after subtracting the corotation contribution
\begin{equation}
\boldsymbol{J}- \rho_e \Omega \rho \hat{\boldsymbol{\phi}}=\Lambda \boldsymbol{B}
\end{equation}
Our definition $\Lambda \equiv (\mathbf{J}\cdot\mathbf{B})/B^{2}$ in equation \eqref{eq:Lambda_def}, together with the decomposition in equation \eqref{eq:JinLambda}, is consistent with this statement, equation~\eqref{eq:JinLambda} separates the field-aligned contribution $\Lambda\,\mathbf{B}$ from the perpendicular drift current $\rho_e\,(\mathbf{E}\times\mathbf{B})/B^{2}$, and in the near-zone corotating magnetosphere the drift term is $O(\epsilon)$.
}

In the near-zone corotating magnetosphere, the electric field is parametrically small compared to the magnetic field: $|\mathbf{E}|/|\mathbf{B}|\sim \epsilon$, where
\begin{equation}
\epsilon \equiv \frac{\Omega R_{\star}}{c} \;=\; \frac{R_\star}{R_{\rm LC}} \ll 1,
\end{equation}
with $\Omega$ the stellar angular velocity, $R_{\star}$ the stellar radius, $c$ the speed of light, and $R_{\rm LC}=c/\Omega$ the light-cylinder radius. Thus the drift term in \eqref{eq:JinLambda} is suppressed by $\mathcal{O}(\epsilon)$, and at leading order we may write
\begin{equation}
\mathbf{J}=\Lambda \mathbf{B}+\mathcal{O}(\epsilon).
\label{eq:J_LO}
\end{equation}
Current conservation, together with stationarity in the corotating frame, implies that the leading-order constraint is the spatial one,
\begin{equation}
0 \;=\; \nabla \cdot \mathbf{J} \;+\; \mathcal{O}(\epsilon)\,.
\end{equation}
{Using equation \eqref{eq:J_LO} together with the $\nabla\cdot\mathbf{B}=0$, one immediately finds, to leading order in $\epsilon$, that $\Lambda$ is constant along magnetic field lines, and therefore can be written as $\Lambda=\Lambda(\alpha,\beta)$ in terms of the Euler potentials $(\alpha,\beta)$ labeling the open flux tubes.}

Field-line constancy alone does not determine the distribution of $\Lambda(\alpha,\beta)$ across the polar cap. The leading-order force-free constraint implies only that $\Lambda$ is constant along field lines, i.e.\ $\Lambda=\Lambda(\alpha,\beta)$, but it leaves the \emph{cross-cap} distribution underdetermined. To proceed one must supply additional physical input connecting the smooth open-cap interior to the global current circuit.
Following the matched-asymptotic closure picture of \citet{Gralla2017} and \citet{Lockhart2019}, we assume that
(i) the open-cap interior is smooth and contains no resolved current sources or sinks at the order retained, and
(ii) global current closure is achieved by a thin, unresolved return-current layer localized on the separatrix, which
supplies boundary data for the smooth interior continuation. {These assumptions motivate a smooth interior boundary-value problem but do not uniquely select $\Lambda(\alpha,\beta)$.}

{In this work we therefore introduce an explicit additional interior closure ansatz:
$\Phi\equiv\Lambda(\alpha,\beta)$ is chosen as the weighted harmonic extension of its separatrix boundary data
(see Appendix~B for derivation and motivation).} Specifically, we require that the smooth interior admit a local surface flux
$F^a$ with $\nabla_a F^a=0$, and we adopt the minimal local and isotropic constitutive relation
$F^a=-\mathcal{W}g^{ab}\nabla_b\Phi$ for $\Phi\equiv\Lambda(\alpha,\beta)$, where $\mathcal{W}>0$ is fixed (up to an
overall normalization) by the natural flux-tube measure on the cap. This yields the following local divergence-form
interior equation (derived in Appendix~B):
\begin{equation}
\nabla_{\perp}\!\cdot\!\bigl(\mathcal{W}\,\nabla_{\perp}\Phi\bigr) \;=\; 0\,,
\label{eq:conservation_W}
\end{equation}
where $\nabla_\perp$ is the covariant derivative intrinsic to the stellar surface. We emphasize that
equation~\eqref{eq:conservation_W} is a \emph{second-order elliptic} constraint governing cross-field variation across the cap.

Here $\mathcal{W}$ is a strictly positive weight encoding the local surface density of open flux tubes and is
proportional to the normal magnetic field component,
\begin{equation}
\mathcal{W} \;\propto\; B_{n} \;\propto\; \frac{1}{\sqrt{g}}\,\epsilon^{ab}\,\partial_{a}\alpha\,\partial_{b}\beta\,,
\end{equation}
{Within a gradient expansion on the cap, equation~\eqref{eq:conservation_W} is the minimal leading-order choice in the class of local, isotropic, self-adjoint second-order divergence-form closures. More general choices would require a tensor coefficient $K^{ab}(x)$ or higher-derivative terms, introducing additional structure and free functions beyond the scope of this paper. For clarity, this differs from the surface-current closure problem in \citet{Beskin2010}, where one solves for surface currents given a prescribed inflowing $\Lambda(\alpha,\beta)$}, $\epsilon^{ab}$ is the Levi--Civita tensor density on the surface and $\partial_{a}$ denotes partial derivatives with respect to surface coordinates. In the small-cap limit one may treat $\mathcal{W}$ as constant to leading order and approximate the cap by a flat unit disk with polar coordinates $(r,\beta)$, where $r\in[0,1)$ labels flux surfaces and $\beta$ is the magnetic azimuth on the surface. We utilize the gauge freedom in the Euler potentials to identify the second potential with the geometric azimuth on the stellar surface. Consequently, we denote both the field-line label and the azimuthal coordinate by $\beta$; while formally a slight abuse of notation, this identification is consistent within the adopted gauge and simplifies the harmonic analysis. Equation~\eqref{eq:conservation_W} then reduces to the flat Laplace equation on the disk, \begin{equation} \left(\partial_r^2+\frac{1}{r}\partial_r+\frac{1}{r^2}\partial_\beta^2\right)\Lambda \;=\;0, \qquad 0\le r<1, 
\label{eq:Laplace_disk} 
\end{equation} whose most general solution that is smooth at $r=0$ is 
\begin{equation} \Lambda(r,\beta)=a_0+\sum_{m=1}^{\infty} r^{m}\Big[a_m\cos\!\big(m(\beta-\beta_m)\big)\Big], \label{eq:harmonic_general} \end{equation}
(with real amplitudes $a_m$ and phases $\beta_m$; sine components are absorbed into $\beta_m$). This step is fully determined by first principles: it follows only from the interior elliptic equation and regularity at the pole.

For an oblique rotator, axial symmetry is broken at order $\sin\iota$, so the leading non-axisymmetric contribution is the $m=1$ harmonic. Truncating \eqref{eq:harmonic_general} to $(m=0,1)$ gives the \emph{unique} minimal smooth interior form,
\begin{equation}
\Lambda(r,\beta)\;\simeq\; A_0 \;+\; A_1\, r\,\cos(\beta-\beta_0),
\label{eq:rigorous_minimal}
\end{equation}
where $A_0\equiv a_0$ and $A_1\equiv a_1$.

For a dipole near the stellar surface, the open-cap flux label is monotonic with magnetic colatitude and one may take, to leading order on the cap,
\begin{equation}
\frac{\alpha}{\alpha_0}=r^2=\sin^2\!\frac{\chi}{2},
\qquad
\chi \equiv 2\arcsin\sqrt{\frac{\alpha}{\alpha_0}}\in[0,\pi],
\label{eq:chi_def}
\end{equation}
so that \eqref{eq:rigorous_minimal} becomes
\begin{equation}
\Lambda(\chi,\beta)\;\simeq\; A_0 \;+\; A_1\,\sin\!\frac{\chi}{2}\,\cos(\beta-\beta_0).
\label{eq:minimal_in_chi}
\end{equation}

Equation~\eqref{eq:minimal_in_chi} is the leading-order \emph{asymptotic} interior result. Beyond this strict limit, higher-order effects (spatial variation of $\mathcal{W}$ across the cap, spherical geometry, and the separatrix boundary layer required for current closure) modify the higher-order coefficients in the small-$\chi$ expansion while preserving: (i) regularity at $\chi=0$, (ii) smoothness in the interior, and (iii) the $m=0$ and $m=1$ symmetry structure.
A natural way to incorporate these higher-order corrections while keeping a compact analytic form is to ``resum'' the radial dependence using smooth single-scale special functions whose Taylor series match the leading asymptotics at the pole.
Specifically,
\begin{equation}
\begin{aligned}
J_0(\chi)&=1-\frac{\chi^2}{4}+\mathcal{O}(\chi^4),\\
J_1(\chi)&=\frac{\chi}{2}-\frac{\chi^3}{16}+\mathcal{O}(\chi^5),
\end{aligned}
\label{eq:Bessel_series}
\end{equation}
while
\begin{equation}
\sin\!\frac{\chi}{2}=\frac{\chi}{2}-\frac{\chi^3}{48}+\mathcal{O}(\chi^5).
\end{equation}
{Thus $J_0(\chi)$ matches the constant interior mode through $\mathcal{O}(\chi^2)$, and $J_1(\chi)$ matches the leading $m=1$ behavior through $\mathcal{O}(\chi)$, with differences entering only at higher order where the strict small-cap Laplace approximation is anyway corrected. Although $\chi$ ranges over $[0,\pi]$ across the full cap, the series expansions above are used only to match the near-axis ($\chi \to 0$) asymptotic behavior that follows from the cap-interior Laplace solution equation~\eqref{eq:minimal_in_chi}. They serve solely to justify that the chosen special-function completion reproduces the correct leading-order coefficients near the pole. This motivates the minimal Bessel-harmonic completion}
\begin{equation}
\begin{aligned}
\Lambda(\chi,\beta)\;&\simeq\;
A_0\,J_0(\chi)\;+\;A_1\,J_1(\chi)\cos(\beta-\beta_0),\\
\chi\;&=2\arcsin\sqrt{\alpha/\alpha_0},
\end{aligned}
\label{eq:Bessel_completion}
\end{equation}
which preserves the first-principles near-axis structure \eqref{eq:minimal_in_chi} but provides a compact global profile on $0\le\chi\le\pi$.
We stress that the step \eqref{eq:minimal_in_chi}\,$\to$\,\eqref{eq:Bessel_completion} is a controlled modeling choice: it is not fixed uniquely by the interior equation alone, but by adopting the simplest single-scale analytic completion consistent with the asymptotics and with the matched-asymptotic picture in which non-smooth behavior resides in the unresolved return-current layer.

The obliquity $\iota$ is defined as the angle between the rotation axis $\boldsymbol{\Omega}$ and the magnetic axis $\mathbf{m}$. Decomposing $\boldsymbol{\Omega}$ into components parallel and perpendicular to $\mathbf{m}$, we write
\begin{equation}
\boldsymbol{\Omega}=\Omega\big(\cos\iota \,\hat{\mathbf{m}} 
\;+\; \sin \iota\, \hat{\mathbf{e}}_1\big),
\end{equation}
where $\hat{\mathbf{e}}_1$ lies in the meridian plane containing $\boldsymbol{\Omega}$ and $\mathbf{m}$. By definition of the magnetic azimuth $\beta$, this meridian corresponds to $\beta=0$, so we set $\beta_0=0$ in \eqref{eq:Bessel_completion}.

Close to the magnetic pole, the leading non-axisymmetric distortion follows the projection of $\boldsymbol{\Omega}$ along the local field direction. Writing
\begin{equation}
\hat{\mathbf{b}}=\cos \theta^{\prime}\,\hat{\mathbf{m}}
\;+\; \sin \theta^{\prime} \big(\cos \beta \,\hat{\mathbf{e}}_1
\;+\; \sin \beta \,\hat{\mathbf{e}}_2\big),
\end{equation}
one finds
\begin{equation}
\boldsymbol{\Omega}\cdot\hat{\mathbf{b}} 
= \Omega\big(\cos \iota \,\cos\theta^{\prime}
+\sin \iota\,\sin \theta^{\prime}\cos\beta\big),
\end{equation}
so the leading $m=1$ distortion is proportional to $\sin\iota\,\cos\beta$. Accordingly, $A_0\propto \cos\iota$ and $A_1\propto \sin\iota$.

The overall normalization is set by far-zone current closure. In the convention of \citet{Gralla2017}, matching to the open-field poloidal current fixes the scale of $\Lambda$ to be $2\Omega$ (up to a sign flip between hemispheres). We therefore write
\begin{equation}
A_0=s\,2\Omega\cos\iota,\qquad A_1=-s\,2\Omega\sin\iota,\qquad \beta_0=0,
\label{eq:obliquity_coeffs}
\end{equation}
where $s=\pm1$ labels the two hemispheres and the relative minus sign is a convention matching \citet{Gralla2017}.

Finally, substituting \eqref{eq:obliquity_coeffs} into \eqref{eq:Bessel_completion} yields the oblique-dipole profile
\begin{equation}
\begin{aligned}
\Lambda(\alpha,\beta)
&= s\,2 \Omega\!\left[
J_0\!\Big(2\arcsin\sqrt{\alpha/\alpha_0}\Big)\cos\iota\right.\\
&\qquad\left.
- J_1\!\Big(2\arcsin\sqrt{\alpha/\alpha_0}\Big)\cos\beta\,\sin\iota
\right],\quad \alpha<\alpha_0\,.
\end{aligned}
\label{eq:our-Lambda}
\end{equation}
Here $J_0$ and $J_1$ are Bessel functions of the first kind of orders zero and one, respectively, and $\alpha<\alpha_0$ specifies that the result applies to field lines anchored within the open-field polar cap.

In summary, steps above are derived directly stationary, ideal force-free fields, and the cap-interior elliptic constraint, yielding the unique minimal smooth interior structure \eqref{eq:rigorous_minimal} . The only additional ingredient is the analytic completion \eqref{eq:Bessel_completion}, justified as the simplest single-scale resummation that preserves the first-principles near-axis asymptotics while providing a compact global profile suitable for far-zone matching, leading to \eqref{eq:our-Lambda}.

\subsection{Axisymmetric quadrupolar field case}
\label{sec:Lambda-quadrupole}

To illustrate the generality of the formulation in Section~\ref{subsec:oblique-dipole}, we now consider a case in which the \emph{far zone} of the magnetosphere is dominated (at leading order) by an $\ell=2$ pattern. In vacuum magnetostatics higher multipoles decay faster with radius than the dipole. Here ``far-zone dominated by $\ell=2$'' should be understood as an idealized force-free configuration sustained by magnetospheric currents, used to illustrate the matching logic.
The near-zone analysis is unchanged: the field-aligned invariant $\Lambda$ satisfies $\mathbf{B}\cdot\nabla\Lambda=0$ and hence may be regarded as a function of the field-line labels $(\alpha,\beta)$. In the small-cap limit the cap-interior (smooth) part of $\Lambda$ obeys the conservation law to leading order (constant $\mathcal{W}$) this reduces to Laplace's equation on the cap interior.

Approximating the cap by a flat unit disk with polar coordinates $(r,\beta)$, the leading-order interior equation is same as \eqref{eq:Laplace_disk}
The most general solution that is smooth at $r=0$ is identical to \eqref{eq:harmonic_general}. Far-zone matching determines which $m$ are needed. An $\ell=2$ pattern decomposes under rotations about the spin axis into azimuthal harmonics with $m=0,1,2$; therefore the minimal smooth cap-interior solution consistent with an $\ell=2$ far-zone source retains precisely these modes:
\begin{equation}
\Lambda(r,\beta)\;\simeq\; A_0 \;+\; A_1\,r\,\cos(\beta-\beta_1)\;+\;A_2\,r^2\,\cos\!\big(2(\beta-\beta_2)\big).
\label{eq:rigorous_minimal_l2}
\end{equation}

Consider a \emph{pure axisymmetric} quadrupole in its principal frame, with principal axis $\hat{\boldsymbol{a}}$ tilted by an angle $\iota_Q$ relative to the rotation axis $\boldsymbol{\Omega}$, and azimuth $\psi_Q$ about $\hat{\boldsymbol{\Omega}}$. In the quadrupole principal frame the only nonzero spherical-tensor component is $Q_{20}$. Under a rotation that carries $\hat{\boldsymbol{a}}\to\hat{\boldsymbol{\Omega}}$, the spin-frame components are
\begin{equation}
Q^{(\Omega)}_{2m}= d^{\,2}_{m0}(\iota_Q)\,e^{-im\psi_Q}\,Q_{20},
\qquad m=0,\pm1,\pm2,
\label{eq:Wigner-rotation}
\end{equation}
with reduced Wigner $d$-matrix elements
\begin{equation}
\begin{aligned}
d^{\,2}_{00}(\iota_Q)&=\tfrac12\big(3\cos^2\iota_Q-1\big),\\
d^{\,2}_{10}(\iota_Q)&=-\sqrt{\tfrac32}\sin\iota_Q\cos\iota_Q,\\
d^{\,2}_{20}(\iota_Q)&=\sqrt{\tfrac38}\sin^2\iota_Q.
\end{aligned}
\label{eq:wigner-d2}
\end{equation}
Because Euler potentials (and hence the labels $(\alpha,\beta)$) are constant along a field sheet in the near zone, the azimuthal dependence carried by $e^{im\beta}$ is transported unchanged along that sheet to the matching region. Therefore, matching to an $\ell=2$ far-zone pattern fixes the cap-side phases to be aligned with $\psi_Q$ and fixes the relative weights of the $m=0,1,2$ harmonics to scale with $d^{2}_{m0}(\iota_Q)$. Concretely, the real $\ell=2$ pattern may be written schematically as
\begin{equation}
\begin{aligned}
&\Re\!\left[\sum_{m=-2}^{2} Q^{(\Omega)}_{2m}\,Y_{2m}(\theta,\phi)\right]
=\;
C_0(\theta)\\
&\quad+\;C_1(\theta)\cos(\phi-\psi_Q)
\;+\;C_2(\theta)\cos\!\big(2(\phi-\psi_Q)\big),
\end{aligned}
\end{equation}
so the corresponding cap-side basis is $\{1,\cos(\beta-\psi_Q),\cos(2(\beta-\psi_Q))\}$, with relative amplitudes
\begin{equation}
C_0:C_1:C_2 \;\propto\; d^{2}_{00}(\iota_Q):d^{2}_{10}(\iota_Q):d^{2}_{20}(\iota_Q),
\label{eq:C_ratios}
\end{equation}
up to an overall normalization and conventions.

Equation~\eqref{eq:rigorous_minimal_l2} is the \emph{rigorous} leading-order small-cap result. Beyond this strict limit, spherical geometry, spatial variation of $\mathcal{W}$, and the separatrix boundary layer modify higher-order coefficients while preserving smoothness at $\chi=0$ and the $(m=0,1,2)$ azimuthal content. As in the dipole case, we adopt a compact single-scale analytic completion by replacing the strict $r^m$ radial factors with smooth functions of $\chi$ that reproduce the correct near-axis scaling. Using
\begin{equation}
\begin{aligned}
&r=\sin\!\frac{\chi}{2}=\frac{\chi}{2}+\mathcal{O}(\chi^3),\\
&J_1(\chi)=\frac{\chi}{2}+\mathcal{O}(\chi^3),\\
&J_2(\chi)=\frac{\chi^2}{8}+\mathcal{O}(\chi^4),
\end{aligned}
\end{equation}
one sees that $r\sim J_1(\chi)$ and $r^2\sim 2J_2(\chi)$ at leading order near the pole. This motivates the minimal completion
\begin{equation}
r \ \mapsto\ J_1(\chi),
\qquad
r^2 \ \mapsto\ 2J_2(\chi),
\label{eq:completion_rules_l2}
\end{equation}
which preserves the first-principles near-axis behavior while providing a compact global profile on $0\le\chi\le\pi$.

Collecting the far-zone weights \eqref{eq:C_ratios}, the hemisphere flip $s=\pm1$, and the overall near--far normalization into a single constant $\mathcal{A}_2$, we obtain the cap current invariant for an axisymmetric tilted quadrupole,
\begin{equation}
\begin{aligned}
\Lambda_{(\text{axisym.\ quad.})}(\alpha,\beta)
&=
s\,\mathcal{A}_{2}\Big[
d^{2}_{00}(\iota_Q)\,J_0(\chi)\\
&+\;d^{2}_{10}(\iota_Q)\,J_1(\chi)\cos(\beta-\psi_Q)\\
&+\;2\,d^{2}_{20}(\iota_Q)\,J_2(\chi)\cos\!\big(2(\beta-\psi_Q)\big)
\Big],
\end{aligned}
\label{eq:Lambda-axisym-quad}
\end{equation}
where $\chi=2\arcsin\sqrt{\alpha/\alpha_0}$ and $\alpha<\alpha_0$.
The sign of $d^{2}_{10}$ may be absorbed into a shift $\psi_Q\to\psi_Q+\pi$ if desired, we keep the standard convention \eqref{eq:wigner-d2}.

For an aligned quadrupole, $\iota_Q=0$ implies $d^{2}_{10}=d^{2}_{20}=0$, so $\Lambda\propto J_0(\chi)$ is axisymmetric, as expected. For small obliquity, $\iota_Q\ll 1$, one has $d^{2}_{10}\propto \iota_Q$ and $d^{2}_{20}\propto \iota_Q^2$, so the leading non-axisymmetric correction is the $m=1$ term $\propto \iota_Q\,J_1(\chi)\cos(\beta-\psi_Q)$, with the $m=2$ term entering at quadratic order.

\subsection{Non-axisymmetric quadrupole}
\label{sec:arbitrary-STF}

The axisymmetric quadrupole considered in Section~\ref{sec:Lambda-quadrupole} is an idealized limiting case. In realistic neutron-star magnetospheres there is no guarantee that the quadrupolar component is axisymmetric, and multiwavelength modeling can favor significantly non-axisymmetric configurations \citep[e.g.][]{Chen_2020}. At the same time, for a \emph{general} non-axisymmetric quadrupole the Euler potentials $(\alpha,\beta)$ are not available in a simple closed form, so constructing the mapping between surface coordinates and $(\alpha,\beta)$ typically requires numerical field-line tracing. It is therefore useful to state the cap current invariant $\Lambda$ in a form that is \emph{conditional on} knowing (i) the quadrupole multipole components in the spin frame and (ii) the cap coordinates $(\chi,\beta)$ assigned to each open field line. Such a result can be evaluated immediately once a numerical map $(\alpha,\beta)\leftrightarrow(\chi,\beta)$ is available, or when an approximate representation of the near-zone flux coordinates becomes available.

We describe a general vacuum quadrupole by a real symmetric trace-free tensor $Q_{ij}$ (five independent components). Let $Q^{(\Omega)}_{2m}$ denote its complex spherical-tensor components in the spin frame aligned with the rotation axis $\boldsymbol{\Omega}$, with $m=-2,-1,0,1,2$. Reality of $Q_{ij}$ implies the standard relation
\begin{equation}
Q^{(\Omega)}_{2,-m} = (-1)^m \big(Q^{(\Omega)}_{2m}\big)^{*},
\label{eq:Q_reality}
\end{equation}
so that $Q^{(\Omega)}_{20}$ is real while the $m=1,2$ components may be complex.

On the polar cap interior, the leading-order conservation law reduces to Laplace's equation in the small-cap limit (constant $\mathcal{W}$), as in \eqref{eq:Laplace_disk}. Approximating the cap by a flat unit disk with polar coordinates $(r,\beta)$, smoothness at $r=0$ implies that each azimuthal harmonic has the minimal regular radial behavior $r^{|m|}$. For an $\ell=2$ far-zone source, the minimal smooth cap-interior solution retains precisely the $|m|\le 2$ modes, which may be written in a compact complex form as
\begin{equation}
\Lambda(r,\beta)\;\simeq\;
C_0
+\sum_{m=1}^{2} r^{m}\Big[
C_m\,e^{\mathrm{i}m\beta}
+(-1)^m C_m^{*}\,e^{-\mathrm{i}m\beta}
\Big].
\label{eq:Lambda_l2_disk_complex}
\end{equation}
This expression is manifestly real and is equivalent to the real cosine form used in \eqref{eq:rigorous_minimal_l2}. Here we adopt the same cap-coordinate convention as above: $\beta$ is the magnetic azimuthal label on the cap (equal to the second Euler potential on the surface by a gauge choice), and $\chi=2\arcsin\sqrt{\alpha/\alpha_0}$ is the cap radial coordinate.

As in Section~\ref{sec:Lambda-quadrupole}, we adopt the same single-scale analytic completion away from the strict $r\ll 1$ limit by replacing the near-axis factors $r^m$ with smooth functions of $\chi$ that reproduce the correct leading behavior. Concretely, we use \eqref{eq:completion_rules_l2} and the associated $m=0$ completion,
\begin{equation}
1\ \mapsto\ J_0(\chi),\qquad
r\ \mapsto\ J_1(\chi),\qquad
r^2\ \mapsto\ 2J_2(\chi),
\label{eq:completion_rules_l2_with_m0}
\end{equation}
which preserves the rigorous near-axis scalings while providing a compact global profile on $0\le\chi\le\pi$.

We now impose rotational covariance and leading-order linearity in the $\ell=2$ matching problem. Under a spatial rotation $R$, the quadrupole multiplet transforms in the $\ell=2$ representation,
\begin{equation}
Q^{(\Omega)\prime}_{2m} = \sum_{m'=-2}^{2} D^{(2)}_{m m'}(R)\,Q^{(\Omega)}_{2m'} ,
\label{eq:Q_Dmatrix}
\end{equation}
where $D^{(2)}_{m m'}(R)$ are transformation matrix elements of $D$-matrix. Since $\Lambda$ is fixed by matching to the far-zone solution sourced by the same multiplet, the corresponding cap amplitudes must transform in the \emph{same} representation. Assuming the leading cap solution depends \emph{linearly} on the far-zone $\ell=2$ source (i.e.\ no mixing with other multipoles at this order), it implies that any rotationally covariant linear map between two equivalent irreducible $\ell=2$ representations is proportional to the identity. Therefore the cap amplitudes are proportional to the spin-frame quadrupole components,
\begin{equation}
C_m=\mathcal{A}_2\,Q^{(\Omega)}_{2m}\quad (m=0,1,2),
\label{eq:C_proportional_Q}
\end{equation}
with a single real constant $\mathcal{A}_2$ fixed by near--far matching and the open-flux normalization $\alpha_0$.

Substituting \eqref{eq:C_proportional_Q} into \eqref{eq:Lambda_l2_disk_complex} and applying \eqref{eq:completion_rules_l2_with_m0} yields the minimal, regular, linear, rotationally covariant cap-interior solution for a general (non-axisymmetric) quadrupole:
\begin{equation}
\begin{aligned}
\Lambda&(\chi,\beta)
=
\mathcal{A}_2\Bigg[
Q^{(\Omega)}_{20}\,J_0(\chi)\\
&+\sum_{m=1}^{2}\mathcal{J}_m(\chi)\Big(
Q^{(\Omega)}_{2m}\,e^{\mathrm{i}m\beta}
+(-1)^m \big(Q^{(\Omega)}_{2m}\big)^{*}\,e^{-\mathrm{i}m\beta}
\Big)
\Bigg],
\end{aligned}
\label{eq:Lambda_Q_complex}
\end{equation}
where we defined $\mathcal{J}_1(\chi)\equiv J_1(\chi)$ and $\mathcal{J}_2(\chi)\equiv 2J_2(\chi)$ to incorporate the same completion used in \eqref{eq:Lambda-axisym-quad}. Using \eqref{eq:Q_reality}, the result may be written explicitly in a real phase-amplitude form. Writing $Q^{(\Omega)}_{2m}=|Q^{(\Omega)}_{2m}|\,e^{\mathrm{i}\varphi_m}$ for $m=1,2$ (with $\varphi_m=\arg Q^{(\Omega)}_{2m}$) and recalling that $Q^{(\Omega)}_{20}$ is real, we obtain
\begin{equation}
\begin{aligned}
\Lambda(\chi,\beta)
&=
\mathcal{A}_2\Big[
Q^{(\Omega)}_{20}\,J_0(\chi)\\
&\quad+\;|Q^{(\Omega)}_{21}|\,J_1(\chi)\cos\!\big(\beta-\varphi_1\big)\\
&\;+\;2\,|Q^{(\Omega)}_{22}|\,J_2(\chi)\cos\!\big(2(\beta-\varphi_2)\big)
\Big].
\end{aligned}
\label{eq:Lambda_Q_real}
\end{equation}
Finally, as in the dipole and axisymmetric-quadrupole cases, the direction of the field-aligned current flips between the two polar caps. Introducing the hemisphere label $s=\pm1$ and restoring the flux-coordinate dependence $\chi=2\arcsin\sqrt{\alpha/\alpha_0}$, we write
\begin{equation}
\begin{aligned}
\Lambda_{(\text{pure }Q)}(\alpha,\beta)
&=
s\,\mathcal{A}_2\Big[
Q^{(\Omega)}_{20}\,J_0(\chi)\\
&\quad+\;|Q^{(\Omega)}_{21}|\,J_1(\chi)\cos\!\big(\beta-\arg Q^{(\Omega)}_{21}\big)\\
&\;+\;2\,|Q^{(\Omega)}_{22}|\,J_2(\chi)\cos\!\big(2(\beta-\arg Q^{(\Omega)}_{22})\big)
\Big],
\end{aligned}
\label{eq:Lambda-arbitrary-quad}
\end{equation}
valid for $\alpha<\alpha_0$. Here $\arg Q^{(\Omega)}_{2m}$ denotes the complex \emph{phase} (argument) of the spin-frame quadrupole coefficient $Q^{(\Omega)}_{2m}$. The appearance of $\arg Q^{(\Omega)}_{2m}$ in \eqref{eq:Lambda-arbitrary-quad} simply encodes the azimuthal orientation of each $m$-harmonic on the polar cap relative to the chosen $\beta=0$ boundary. All quadrupole geometry is encoded in the spin-frame components $Q^{(\Omega)}_{2m}$, obtained by rotating any chosen parametrization of the quadrupole moment tensor $Q_{ij}$, while the radial dependence is fixed by the cap-interior harmonic basis through $J_0,J_1,J_2$ and the completion \eqref{eq:completion_rules_l2_with_m0}. The single amplitude $\mathcal{A}_2$ is determined by matching and normalization choices once $\alpha_0$ is specified.

Inserting the axisymmetric-tilted multiplet \eqref{eq:Wigner-rotation} into \eqref{eq:Lambda_Q_real} (and absorbing the overall $Q_{20}$ amplitude into $\mathcal{A}_2$) reproduces the axisymmetric expression \eqref{eq:Lambda-axisym-quad}, including the same $m=2$ completion factor.

\begin{figure*}[t]
\centering
\includegraphics[width=0.9\textwidth]{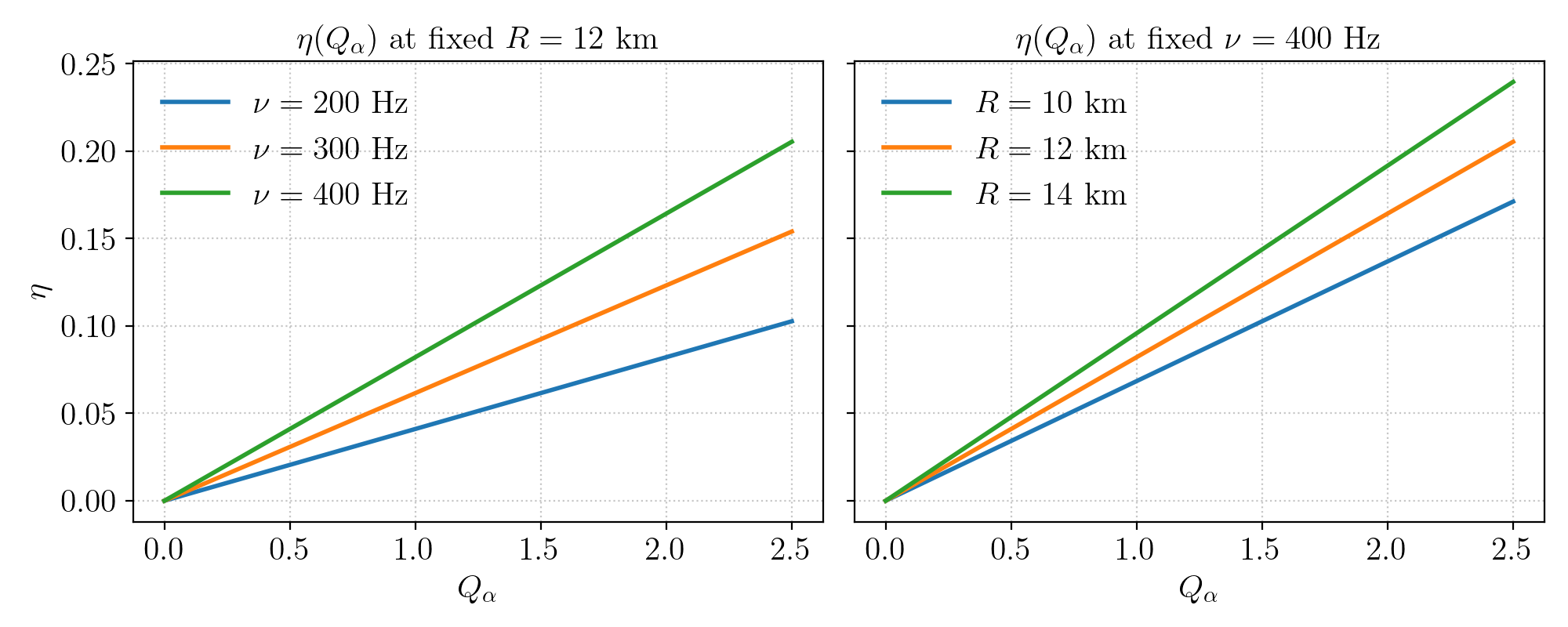}
\caption{
Mapping between the near-zone quadrupole amplitude $Q_\alpha$ and the far-zone dipole--quadrupole mixing parameter
$\eta \equiv \left(B_Q/B_D\right)_{\rm RMS}(R_{\rm LC})$ (evaluated using equation~\eqref{eq:eta-Qalpha-lockhart}).
Left: $\eta(Q_\alpha)$ at fixed stellar radius $R=12\,{\rm km}$ for three spin frequencies $\nu=200,300,400\,{\rm Hz}$.
Right: $\eta(Q_\alpha)$ at fixed $\nu=400\,{\rm Hz}$ for three radii $R=10,12,14\,{\rm km}$.
The representative sequence used in Figures~\ref{fig:jsq-maps}--\ref{fig:erpp-nsx} corresponds to $Q_\alpha=0.50,1.00,1.50$
(i.e., $\eta\simeq 0.041,\,0.082,\,0.123$ for $\nu=400\,{\rm Hz}$ and $R=12\,{\rm km}$).
}
\label{fig:qlambda}
\end{figure*}

\subsection{Dipole--Quadrupole Far--Zone Mixing}
\label{sec:Lambda-DplusQ}

Although Sections~\ref{sec:Lambda-quadrupole}--\ref{sec:arbitrary-STF} treated a (formally) quadrupole-dominated far zone to motivate the analytic construction, it is useful to assess when a quadrupolar contribution can be dynamically relevant at large radii. Outside the star, vacuum multipoles decay with different radial scalings: a multipole of degree $\ell$ falls as $B_\ell\propto r^{-(\ell+2)}$. Therefore, at $r\sim R_{\rm LC}\equiv c/\Omega$ (the light-cylinder radius), the quadrupole-to-dipole ratio scales as
\begin{equation}
\begin{aligned}
\frac{B_{Q}}{B_{D}}\Big|_{R_{\rm LC}}
&\;\simeq\;
\Big(\frac{B_{Q}}{B_{D}}\Big)_{R_\star}\,
\frac{R_\star}{R_{\rm LC}}
\;=\;
\Big(\frac{B_{Q}}{B_{D}}\Big)_{R_\star}\,
\epsilon,\\
\text{where}\quad\epsilon
&\equiv
\frac{\Omega R_\star}{c}=\frac{R_\star}{R_{\rm LC}}.
\end{aligned}
\label{eq:BQBD-LC}
\end{equation}
Here $B_Q$ and $B_D$ denote characteristic quadrupolar and dipolar field strengths, and $R_\star$ is the stellar radius. For ordinary pulsars with $P\sim 0.1$--$1\,$s one has $\epsilon\sim 10^{-3}$--$10^{-4}$, whereas for MSPs with $P\sim 1$--$5\,$ms one finds $\epsilon\sim 0.04$--$0.25$. Equation~\eqref{eq:BQBD-LC} shows that a \emph{pure} quadrupolar far zone would require $(B_Q/B_D)_{R_\star}\gg \epsilon^{-1}$, i.e.\ an extremely quadrupole-dominated surface field. This appears disfavored in typical systems: while crustal Hall evolution, strong internal currents, or accretion can generate substantial near-surface multipoles, the steeper radial falloff generically leaves the far zone dipole-dominated \citep{Goldreich1992,Gourgouliatos2014,Geppert2014,Ciolfi2013,Braithwaite2006,Payne2004,petri2015}.

A quadrupole can nevertheless contribute non-negligibly at $R_{\rm LC}$ in some regimes, particularly for MSPs. As a practical diagnostic, we define the quadrupolar \emph{fraction} of the spin-down power by
\begin{equation}
f_Q \;\equiv\; \frac{L_Q}{L_D+L_Q}
\;\approx\;
\frac{B_Q^2}{B_D^2+B_Q^2}\Big|_{R_{\rm LC}},
\label{eq:fQ-def}
\end{equation}
where $L_D$ and $L_Q$ denote the dipolar and quadrupolar contributions to the Poynting flux. Equation~\eqref{eq:fQ-def} should be read as an order-of-magnitude proxy: in a fully force-free magnetosphere, the partition of the Poynting flux into multipolar ``luminosities'' can differ from the vacuum scaling by order-unity factors. For our purposes, $f_Q$ provides a convenient measure of the relative far-zone strength.
Introducing the far-zone amplitude ratio
\begin{equation}
\eta \;\equiv\; \frac{B_Q}{B_D}\Big|_{R_{\rm LC}}
\approx \Big(\frac{B_Q}{B_D}\Big)_{R_\star}\epsilon,
\qquad
f_Q \;=\; \frac{\eta^2}{1+\eta^2},
\label{eq:eta-def}
\end{equation}
we note that $f_Q=0.2$ corresponds to $\eta\simeq 0.5$. Combining this with equation~\eqref{eq:BQBD-LC} implies that achieving $f_Q\gtrsim 0.2$ requires $(B_Q/B_D)_{R_\star}\sim \eta/\epsilon$, i.e.\ $(B_Q/B_D)_{R_\star}\sim \mathcal{O}(10^2)$ for a $P\sim 0.1\,$s pulsar but only $(B_Q/B_D)_{R_\star}\sim \mathcal{O}(1$--$10)$ for MSPs. This motivates focusing on the mixed dipole--quadrupole case, especially for MSPs.

Given these considerations, the most relevant and testable scenario is a mixed far zone in which the dipole dominates but the quadrupole contributes at the level $\eta\sim 0.1$--$1$. At the level of the leading-order matched-asymptotic construction used throughout this paper, the cap-interior problem is linear and the dependence on the far-zone source enters through matching. In the minimal approximation, this implies that a dipole--quadrupole far zone leads to a cap invariant $\Lambda$ given by a linear superposition of the dipole and quadrupole templates derived in Sections~\ref{subsec:oblique-dipole} and~\ref{sec:Lambda-quadrupole}. For concreteness, and because our current Euler-potential pipeline is most mature for the axisymmetric ($m=0$) quadrupole family, we present the mixed solution obtained by adding the oblique-dipole expression \eqref{eq:our-Lambda} to the tilted-axisymmetric-quadrupole expression \eqref{eq:Lambda-axisym-quad}:
\begin{equation}
\begin{aligned}
&\Lambda_{(D+Q)}(\alpha,\beta)
={}
s\,2\Omega\Big[J_0(\chi)\cos\iota - J_1(\chi)\cos\beta\,\sin\iota\Big]
\\
&+
s\,\mathcal{A}_2\,\Big[
d^{\,2}_{00}(\iota_Q)\,J_0(\chi)
-2\,|d^{\,2}_{10}(\iota_Q)|\,J_1(\chi)\cos\!\big(\beta-\psi_Q\big)
\\
&
+2\,|d^{\,2}_{20}(\iota_Q)|\,J_2(\chi)\cos\!\big(2(\beta-\psi_Q)\big)
\Big].
\end{aligned}
\label{eq:Lambda-mixed-sum}
\end{equation}
where $s=\pm1$ labels the two hemispheres, $\chi=2\arcsin\sqrt{\alpha/\alpha_0}$, and $\beta$ is the cap azimuthal label. The phase $\psi_Q$ is the azimuth of the quadrupole axis about $\boldsymbol{\Omega}$, measured relative to the dipole meridian that defines $\beta=0$ in the oblique-dipole template. The quantities $d^{2}_{m0}(\iota_Q)$ are matrix elements for $\ell=2$ defined in \eqref{eq:wigner-d2}  in Section~\ref{sec:Lambda-quadrupole}. The absolute values enforce the phase convention used in equation~\eqref{eq:Lambda-axisym-quad}, in which the azimuthal dependence is carried entirely by the cosine factors.

To relate the near-cap amplitude $\mathcal{A}_2$ to the far-zone ratio $\eta$ in equation~\eqref{eq:eta-def}, we adopt the practical normalization
\begin{equation}
\mathcal{A}_2 \;=\; (2\Omega)\,\kappa\,\eta,
\label{eq:A2-cal}
\end{equation}
where the order-unity factor $\kappa$ absorbs geometric differences between the dipole and quadrupole mappings to the open-flux boundary $\alpha_0$ (and, more generally, force-free versus vacuum normalizations). In the numerical examples below we take $\kappa=1$ unless stated otherwise. With this choice, $\eta$ (or equivalently $f_Q$) can be interpreted as a far-zone mixing parameter that controls the relative size of the $\ell=2$ contribution.

Substituting equation~\eqref{eq:A2-cal} into equation~\eqref{eq:Lambda-mixed-sum} yields the working mixed template
\begin{equation}
\begin{aligned}
&\Lambda_{(D+Q)}(\alpha,\beta)
=\;
s\,2\Omega\Big\{
J_0(\chi)\cos\iota
- J_1(\chi)\cos\beta\,\sin\iota \\
&
+\kappa\eta\Big[
d^{\,2}_{00}(\iota_Q)\,J_0(\chi)
-2\,|d^{\,2}_{10}(\iota_Q)|\,J_1(\chi)\cos\!\big(\beta-\psi_Q\big) \\
&
+2\,|d^{\,2}_{20}(\iota_Q)|\,J_2(\chi)\cos\!\big(2(\beta-\psi_Q)\big)
\Big]
\Big\}.
\end{aligned}
\label{eq:Lambda-mixed}
\end{equation}
In our numerical implementation, the mixed near-zone magnetic geometry is specified at the stellar surface via the Euler potential $\alpha$ using the multipolar expansion of \citet{Lockhart2019}. Specializing their construction to axisymmetric $\ell=1,2$ modes about a common magnetic axis (magnetic colatitude $\theta'$), we write
\begin{align}
\alpha(r,\theta')
&= R_\star^{2}\Big[
B_1\,R^{>}_1(r)\,\Theta_1(\theta')
- B_2\,R^{>}_2(r)\,\Theta_2(\theta')
\Big],
\label{eq:alpha-mixed-lockhart}
\end{align}
with angular eigenfunctions
\begin{equation}
\Theta_1(\theta') = \sin^2\theta',
\qquad
\Theta_2(\theta') = \cos\theta'\,\sin^2\theta'.
\end{equation}
The radial eigenfunctions $R^{>}_\ell(r)$ encode GR corrections to the external field and, for $\ell=1,2$, are
\begin{subequations}
\label{eq:R-lockhart}
\begin{align}
R^{>}_1(r) &= \frac{3}{2r}\,
\frac{3 - 4f + f^2 + 2\ln f}{(1-f)^3},
\\
R^{>}_2(r) &= \frac{10}{3r^2}\,
\frac{17 - 9f - 9f^2 + f^3 + 6(1+3f)\ln f}{(1-f)^5},
\end{align}
\end{subequations}
where $f(r)\equiv 1-2M/r$ is the usual redshift factor.

For later convenience we define the near-surface normalizations
\begin{equation}
\mu \;\equiv\; B_1 R_\star^{2} R^{>}_1(R_\star),
\qquad
Q_\alpha \;\equiv\; B_2 R_\star^{2} R^{>}_2(R_\star),
\end{equation}
and dimensionless radial profiles
\begin{equation}
\hat{R}_\ell(r) \;\equiv\; \frac{R^{>}_\ell(r)}{R^{>}_\ell(R_\star)}.
\end{equation}
Equation~\eqref{eq:alpha-mixed-lockhart} can then be written as
\begin{align}
\alpha(r,\theta')
&= \mu\,\hat{R}_1(r)\,\sin^2\theta'
- Q_\alpha\,\hat{R}_2(r)\,\cos\theta'\,\sin^2\theta',
\label{eq:alpha-mu-Q}
\end{align}
so that $\mu$ sets the overall dipole strength while $Q_\alpha$ sets the quadrupole strength in the same units (the sign of $Q_\alpha$ encodes the quadrupole polarity; the magnitude controls the mixing).

We relate the near-zone coefficients $(\mu,Q_\alpha)$ to the far-zone mixing parameter $\eta$ using an angle-averaged (rms) field ratio,
\begin{equation}
\bigg(\frac{B_Q}{B_D}\bigg)_{\rm RMS}(r)
\;\equiv\;
\left[
\frac{\displaystyle\int \! \mathrm{d}\Omega\,
\mathbf{B}_Q^2(r,\theta')}{\displaystyle\int \! \mathrm{d}\Omega\,
\mathbf{B}_D^2(r,\theta')}
\right]^{1/2}.
\end{equation}
Using $\mathbf{B}=\nabla\alpha\times\nabla\beta$ with $\beta=\phi_0$, inserting equation~\eqref{eq:alpha-mu-Q}, and evaluating the angular integrals (Appendix~A) yields, to leading order in $R_\star/r$,
\begin{equation}
\bigg(\frac{B_Q}{B_D}\bigg)_{\rm RMS}(r)
\simeq
\frac{2}{\sqrt{6}}\,
\frac{Q_\alpha}{\mu}\,\frac{R_\star}{r}.
\end{equation}
Evaluating this at $r=R_{\rm LC}=c/\Omega$ gives the working relation
\begin{equation}
\eta \;\equiv\;
\frac{B_Q}{B_D}\bigg|_{R_{\rm LC}}
\simeq
\frac{2}{\sqrt{6}}\,
\frac{Q_\alpha}{\mu}\,\epsilon,
\qquad
\epsilon \equiv \frac{R_\star}{R_{\rm LC}}.
\label{eq:eta-Qalpha-lockhart}
\end{equation}
Any residual order-unity uncertainty in this mapping (e.g.\ from alternative averaging prescriptions or higher-order GR corrections) is absorbed into the geometric factor $\kappa$ in equation~\eqref{eq:A2-cal}. Consequently, once the near-zone field is specified by $(\mu,Q_\alpha)$, the far-zone mixing parameter $\eta$ (and thus $f_Q$) is fixed self-consistently by equation~\eqref{eq:eta-Qalpha-lockhart}.

\subsection{Temperature distribution and pulse-profile computation}
\label{subsec:temp-ppm}

With the mixed-field definition of $\Lambda$ in hand, the next step is to compute the near-zone four-current and construct a hotspot temperature map. Following \citet{Gralla2017}, the spatial components of the four-current in spherical coordinates $(r,\theta,\phi)$ can be written, to leading order in $\epsilon$,
\begin{subequations}\label{eq:Current}
\begin{align}
    J^r&=\frac{\Lambda(\alpha,\beta)}{r^2\sin{\theta}}F_{\theta\phi}^{(0)}
    +\mathcal{O}(\epsilon^2),\\
    J^\theta&=\frac{\Lambda(\alpha,\beta)}{r^2\sin{\theta}}F_{\phi r}^{(0)}
    +\mathcal{O}(\epsilon^2),\\
    J^\phi&=\frac{\Lambda(\alpha,\beta)}{r^2\sin{\theta}}F_{r\theta}^{(0)}
    +\mathcal{O}(\epsilon^2),
\end{align}
\end{subequations}
where $F_{\mu\nu}^{(0)}$ are the components of the background vacuum field-strength tensor $F^{(0)}=d\alpha\wedge d\beta$ constructed from the Euler potentials $\alpha$ and $\beta$, and $\epsilon\equiv\Omega R_\star/c=R_\star/R_{\rm LC}$ is the small rotation parameter. The notation $\mathcal{O}(\epsilon^2)$ denotes higher-order corrections in $\epsilon$ that we neglect here. Equation~\eqref{eq:Current} makes explicit why determining $\Lambda(\alpha,\beta)$ is essential: it fixes the leading-order spatial current on open field lines once the background field $F^{(0)}$ is specified.

The time component of the four-current, $J^t$, encodes the charge density. In the same formalism it can be expressed as
\begin{equation}
\begin{split}
J^t= & \frac{\Omega-\Omega_Z}{r(r-2 M)}\left\{\partial_\theta \alpha \partial_\theta \partial_\phi \beta-\partial_\theta \beta \partial_\theta \partial_\phi \alpha\right. \\
& +r(r-2 M)\left(\partial_r \alpha \partial_r \partial_\phi \beta-\partial_r \beta \partial_r \partial_\phi \alpha\right) \\
& -\partial_\phi \alpha\left[\left(1-\frac{2 M}{r}\right) \partial_r\left(r^2 \partial_r \beta\right)\right.\\
&\left.+\frac{\partial_\theta\left(\sin \theta \partial_\theta \beta\right)}{\sin \theta}\right] +\partial_\phi \beta\left[\left(1-\frac{2 M}{r}\right) \right.\\
&\left.\partial_r\left(r^2 \partial_r \alpha\right)
+\frac{\partial_\theta\left(\sin \theta \partial_\theta \alpha\right)}{\sin \theta}\right]\},
\end{split}
\label{eq:jt}
\end{equation}
where $M$ is the stellar mass, $\Omega_Z$ is the frame-dragging (zero-angular-momentum) angular velocity, and $\partial_r$, $\partial_\theta$, and $\partial_\phi$ denote partial derivatives with respect to $(r,\theta,\phi)$. While $J^t$ depends only on the Euler potentials and the spacetime, the \emph{ratio} of the spatial current to the charge density---and hence the pair-creation regime---is controlled by $\Lambda$ through equation~\eqref{eq:Current}.

Analytic work by \citet{2008ApJ...683L..41B} and numerical simulations by \citet{2013MNRAS.429...20T} indicate that particle acceleration above the polar cap is controlled by the ratio $j/(c\rho_\mathrm{GJ})$, where $j$ is the magnitude of the three-current density and $\rho_\mathrm{GJ}$ is the Goldreich--Julian charge density (equivalently, the orthonormal-frame charge density associated with $J^t$ in equation~\eqref{eq:jt}, up to the usual sign conventions). When this ratio exceeds unity (spacelike four-current, $J^2>0$) or becomes negative (opposite signs of $j$ and $\rho_\mathrm{GJ}$), a large potential drop can develop along the field line, accelerating particles to ultra-relativistic Lorentz factors and triggering copious $e^\pm$ pair creation. A fraction of the resulting pairs returns to the polar cap, depositing energy and heating the stellar surface. In contrast, when $0 < j/(c\rho_\mathrm{GJ}) < 1$ (timelike four-current, $J^2<0$), the plasma can stream along the field lines with only mildly relativistic Lorentz factors, without establishing a large accelerating voltage \citep{2013ApJ...762...76C}.
{
We follow the prescription of \citet{Lockhart2019} to convert the computed return current into a hotspot
temperature distribution. In that approach, the local effective temperature is set by the deposited energy
flux, which scales with the product of return-current density and the energy per returning charge. We therefore
adopt the mapping
\begin{equation}
T(\theta,\phi) \propto \Bigl(|j(\theta,\phi)|\,\Delta\Phi[r_c(\theta,\phi)]\Bigr)^{1/4},
\label{eq:temp_map}
\end{equation}
where $|j|$ is the norm of the three-vector current density $(J^{r},J^{\theta},J^{\phi})$ and
$r_c\in[0,1]$ is the polar-cap radial coordinate  with $r=1$ on the last open field line. We take a simple cap-edge potential-drop profile
\begin{equation}
\Delta\Phi(r) \equiv 1-r_c^{2},
\label{eq:deltaPhi_profile}
\end{equation}
which enforces $\Delta\Phi\to 0$ at the polar-cap boundary.
}

Using the resulting temperature map as input, we compute synthetic X-ray light curves and energy-resolved pulse profiles using the pulse-profile modeling (PPM) framework developed in \citet{GPU_ppm}. In this work we employ the CPU implementation and adopt the standard validated numerical resolution tier described in \citet{GPU_ppm}. To separate the impact of magnetic geometry from emission-model systematics, we present results for both isotropic blackbody emission and the NSX hydrogen atmosphere model \citep{Ho_2001,Ho_2003}. In all cases we include the relevant relativistic effects, which are general-relativistic light bending and gravitational redshift, special-relativistic Doppler boosting and aberration, and spin-induced stellar oblateness, but we simplify the observational pipeline by omitting interstellar absorption and a detailed instrument-response convolution. For further details of the forward-modeling implementation, we refer to \citet{Bogdanov_19,GPU_ppm}.

\begin{table}
  \centering
  \caption{
    Fiducial neutron star and observer setup used for all
    energy-resolved pulse-profile and light-curve calculations.
    Symbols denote the physical quantities used in the text and in the
    light-curve code: $M$ (mass), $R$ (radius), $\nu$ (spin frequency),
    $i$ (observer inclination), $d$ (distance), $T_{\rm exp}$ (exposure
    time), ${\tt Nside}$ (HEALPix surface resolution),
    ${\tt fine\_phase}$ (number of fine phase bins),
    ${\tt points\_per\_bin}$ (integration points per observed phase
    bin), and ${\tt setting}$ (numerical configuration flag).
    The spectral model is chosen to be either blackbody (BB) or NSX,
    as specified for each panel in the figures.
  }
  \label{tab:pp-setup}
  \begin{tabular}{lc}
    \hline
    Quantity       & Value \\
    \hline
    Neutron star mass, $M$        & $1.4\,M_\odot$ \\
    Neutron star radius, $R$      & $12\ \mathrm{km}$ \\
    Spin frequency, $\nu$         & $400\ \mathrm{Hz}$ \\
    Observer inclination, $i$     & $1.2\ \mathrm{rad}$ \\
    Distance, $d$                 & $0.3\ \mathrm{kpc}$ \\
    Exposure time, $T_{\rm exp}$  & $5\times10^5\ \mathrm{s}$ \\
    Surface resolution, ${\tt Nside}$          & $64$ \\
    Fine phase resolution, ${\tt fine\_phase}$ & $32\times 6$ \\
    Integration points per bin, ${\tt points\_per\_bin}$ & $2$ \\
    Numerical configuration, ${\tt setting}$   & {\tt "std"} \\
    Spectral model               & BB or NSX \\
    \hline
  \end{tabular}
\end{table}

\begin{figure*}[t]
\centering
\includegraphics[width=\textwidth]{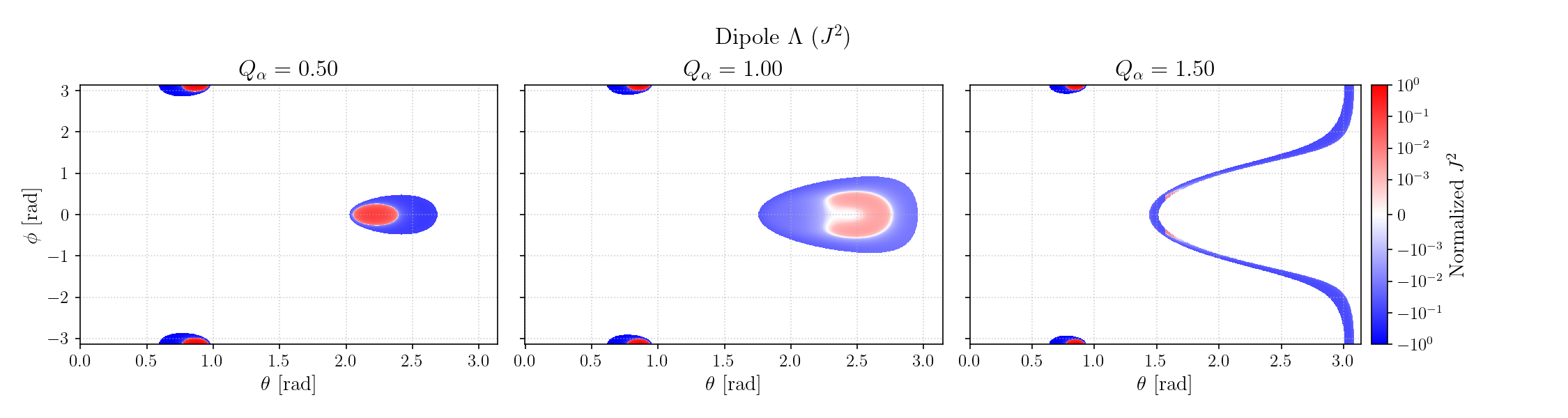}\\
\includegraphics[width=\textwidth]{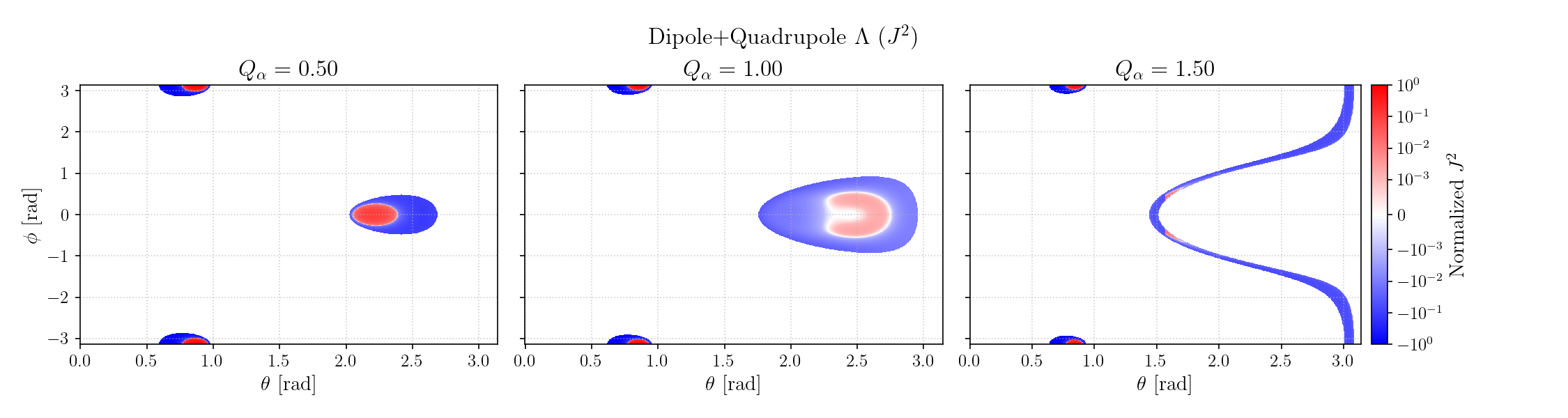}\\
\includegraphics[width=\textwidth]{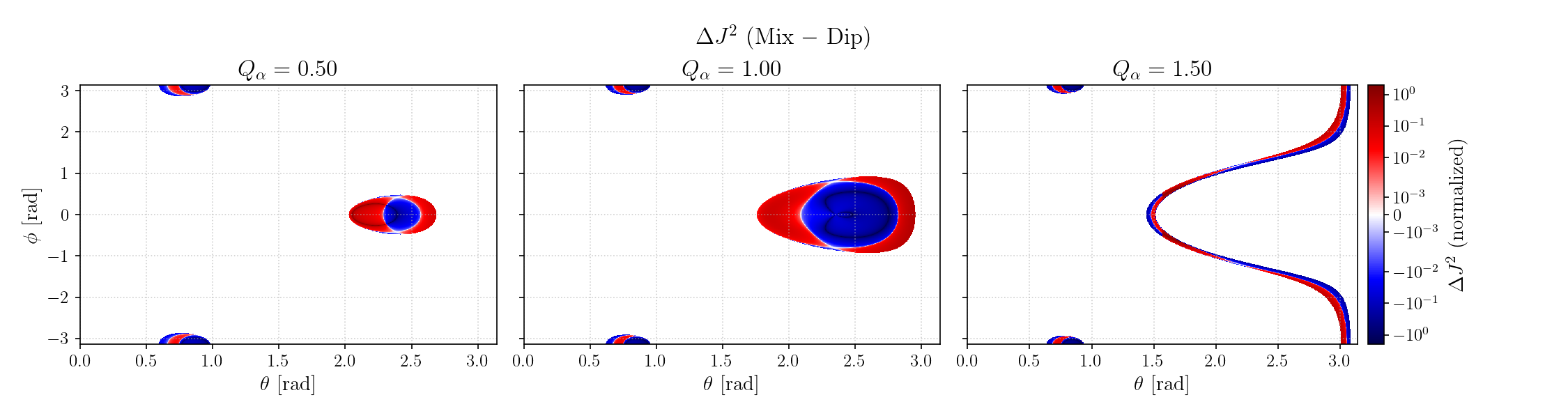}
\caption{
Surface maps of the 4-current invariant $J^2$ in magnetic spherical coordinates $(\theta,\phi)$ for three quadrupole amplitudes
(columns: $Q_\alpha=0.50,1.00,1.50$).
Top row: dipole-only flux-conservation functional $\Lambda$.
Middle row: quadrupole-aware (\dqmix) $\Lambda$.
Bottom row: residual $\Delta J^2 \equiv J^2_{\rm mix}-J^2_{\rm dip}$.
Colors use a symmetric-logarithmic scale and are normalized by the maximum absolute value in each panel; red indicates $J^2>0$
(spacelike 4-current) and blue indicates $J^2<0$ (timelike 4-current).
}
\label{fig:jsq-maps}
\end{figure*}
\begin{figure*}[t]
\centering
\includegraphics[width=\textwidth]{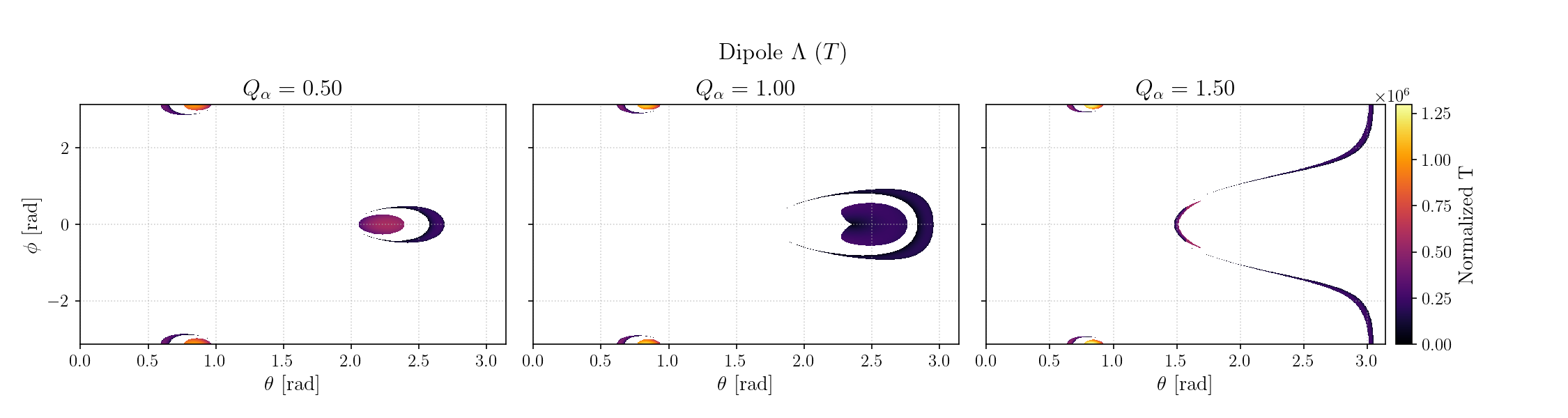}\\
\includegraphics[width=\textwidth]{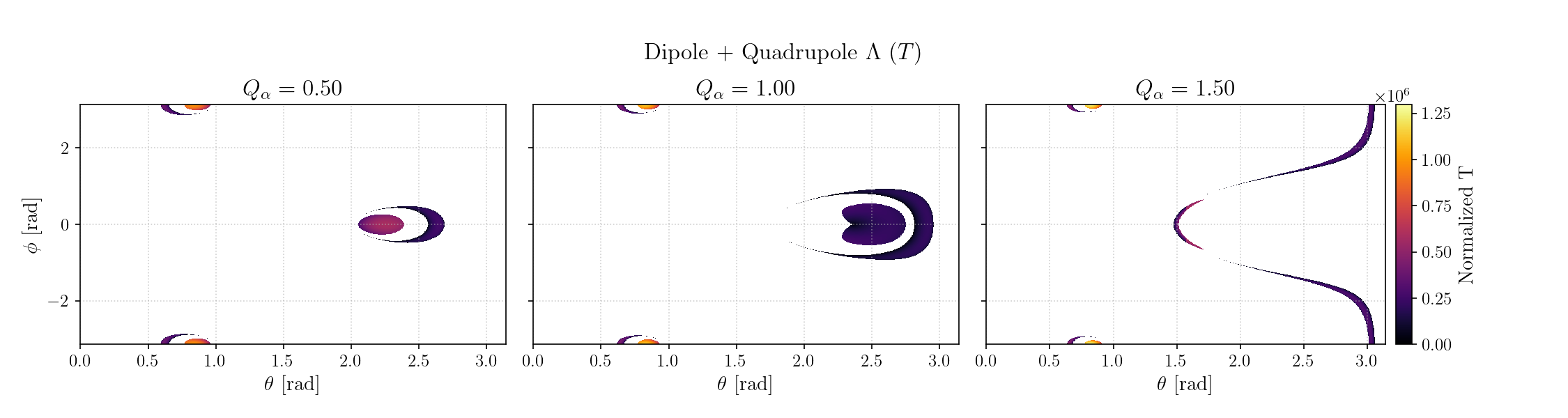}\\
\includegraphics[width=\textwidth]{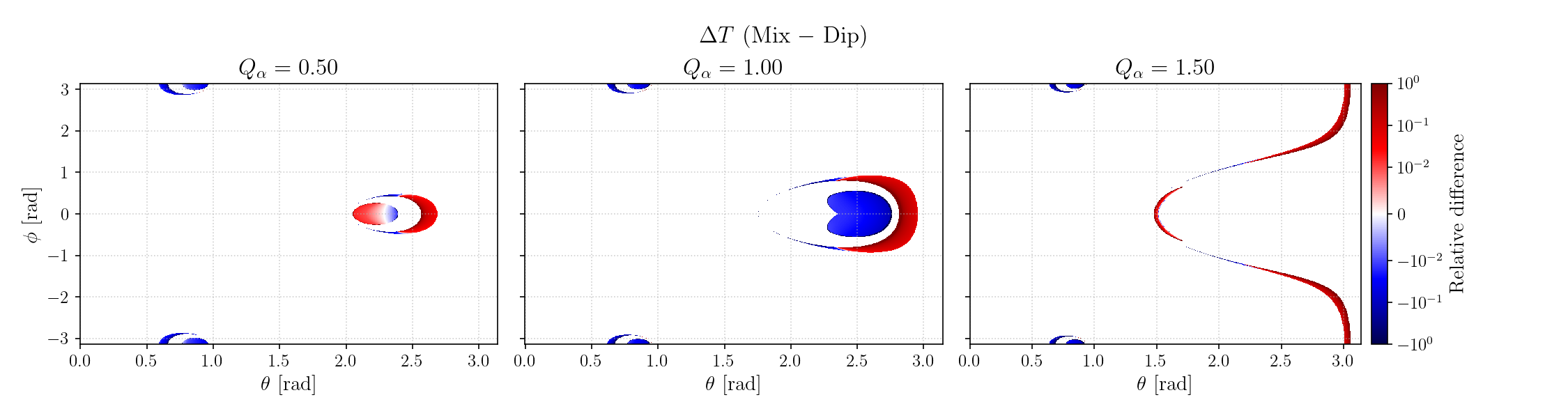}
\caption{Surface effective-temperature patterns inferred from the return-current prescription $T\propto |{\bf j}|^{1/4}$ (overall normalization arbitrary),
shown in magnetic coordinates $(\theta,\phi)$ for the same $Q_\alpha$ sequence as Fig.~\ref{fig:jsq-maps}.
Top row: dipole-only $\Lambda$.
Middle row: \dqmix $\Lambda$.
Bottom row: relative difference $\Delta T \equiv (T_{\rm mix}-T_{\rm dip})/T_{\rm dip}$ (symmetric-logarithmic color scale).
In the top and middle rows, each panel is normalized by its own maximum temperature, i.e.\ colors show $T/T_{\max}$.}
\label{fig:temp-maps}
\end{figure*}

\begin{figure*}[t]
\centering
\includegraphics[width=\textwidth]{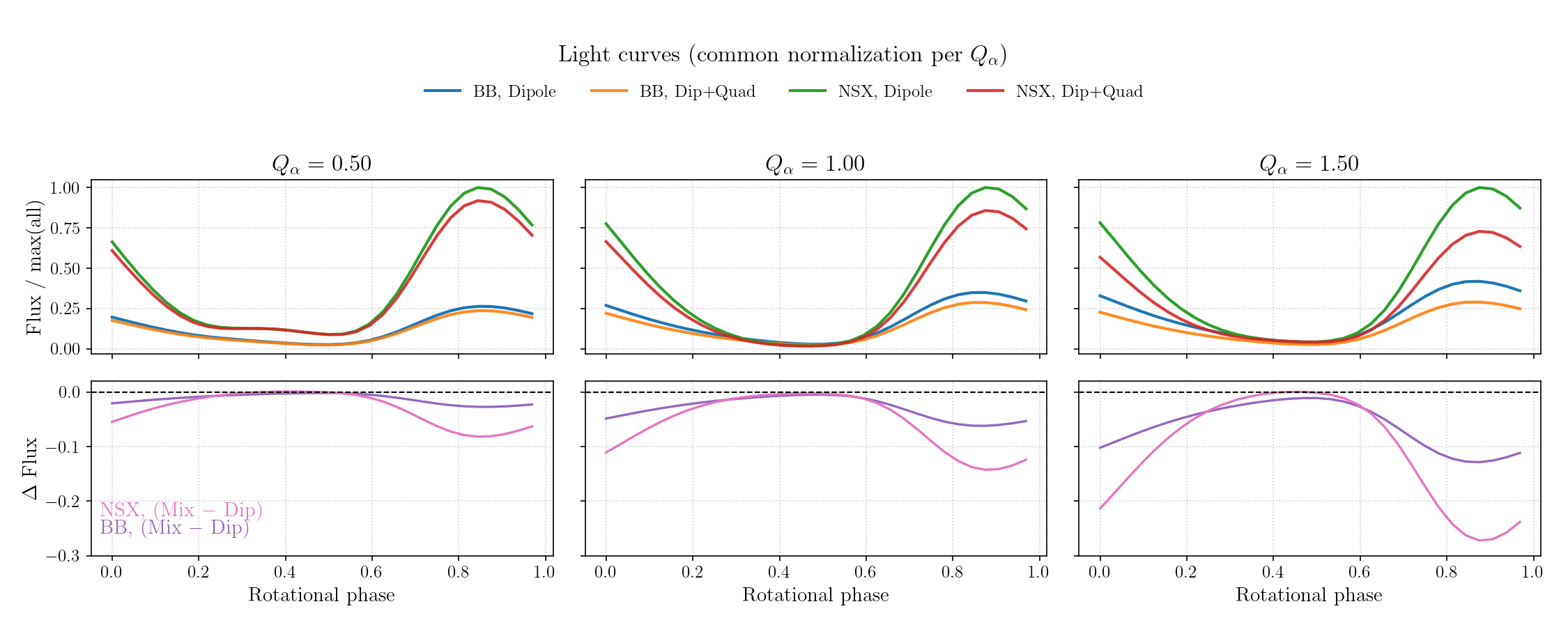}
\caption{Normalized bolometric light curves integrated over $0.1$--$5.2125\,{\rm keV}$ (NICER band) for three quadrupole amplitudes
(columns: $Q_\alpha=0.50,1.00,1.50$), computed for both isotropic blackbody emission (BB) and a beamed NSX atmosphere.
Top row: light curves for the dipole-only and \dqmix $\Lambda$ prescriptions (legend), plotted versus rotational phase.
Within each panel, the flux is normalized by the maximum among the four curves shown to show the relative magnitude between different light curves.
Bottom row: residuals $\Delta F \equiv F_{\rm mix}-F_{\rm dip}$ shown separately for BB and NSX.
All light-curve calculations use the fiducial numerical/geometry setup listed in Table~1.}
\label{fig:lc-panels}
\end{figure*}

\begin{figure*}[t]
\centering
\includegraphics[width=\textwidth]{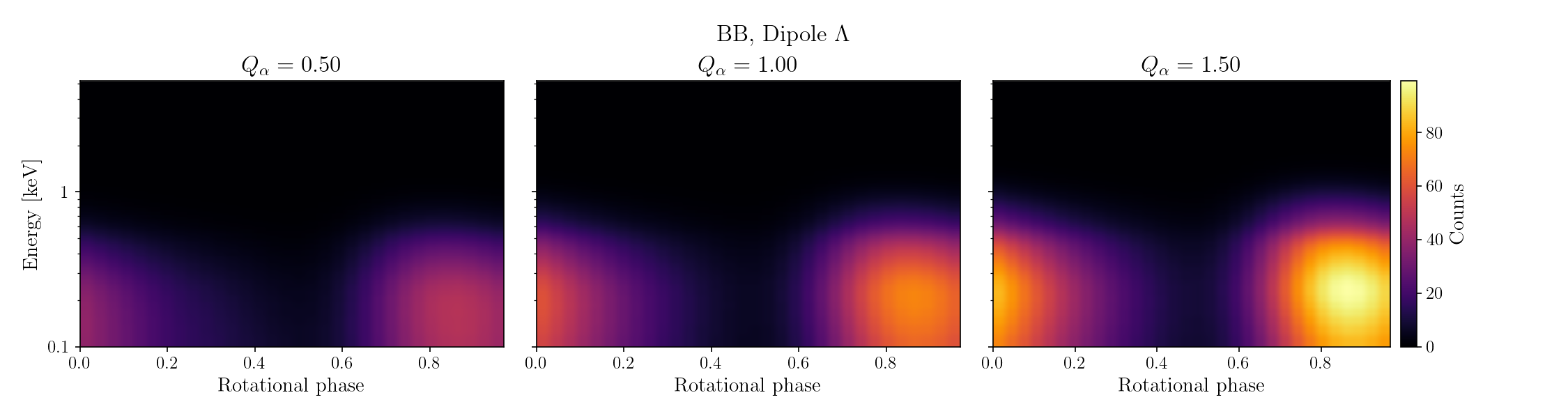}\\
\includegraphics[width=\textwidth]{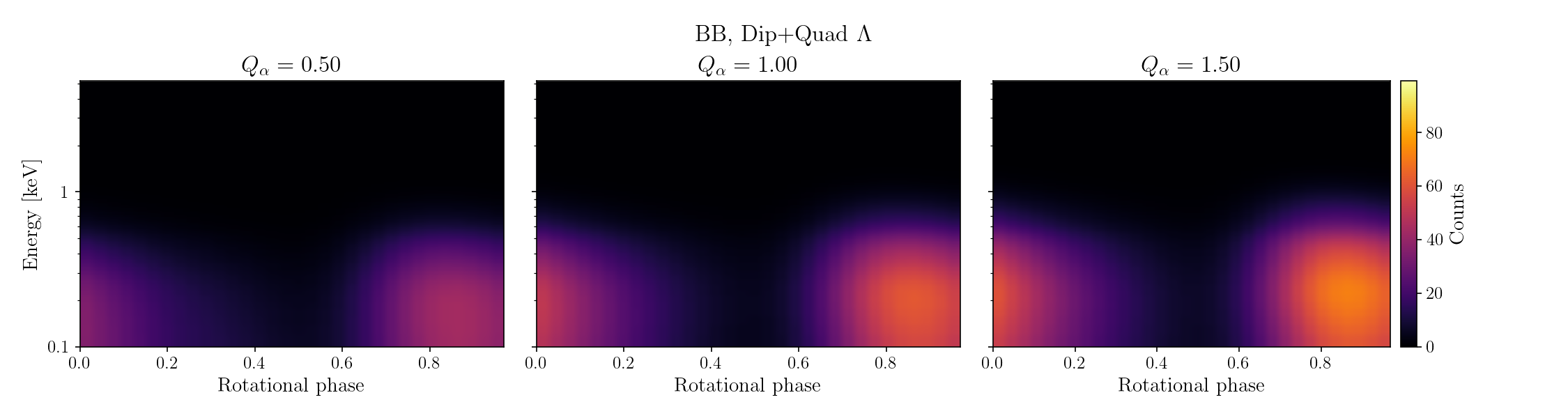}\\
\includegraphics[width=\textwidth]{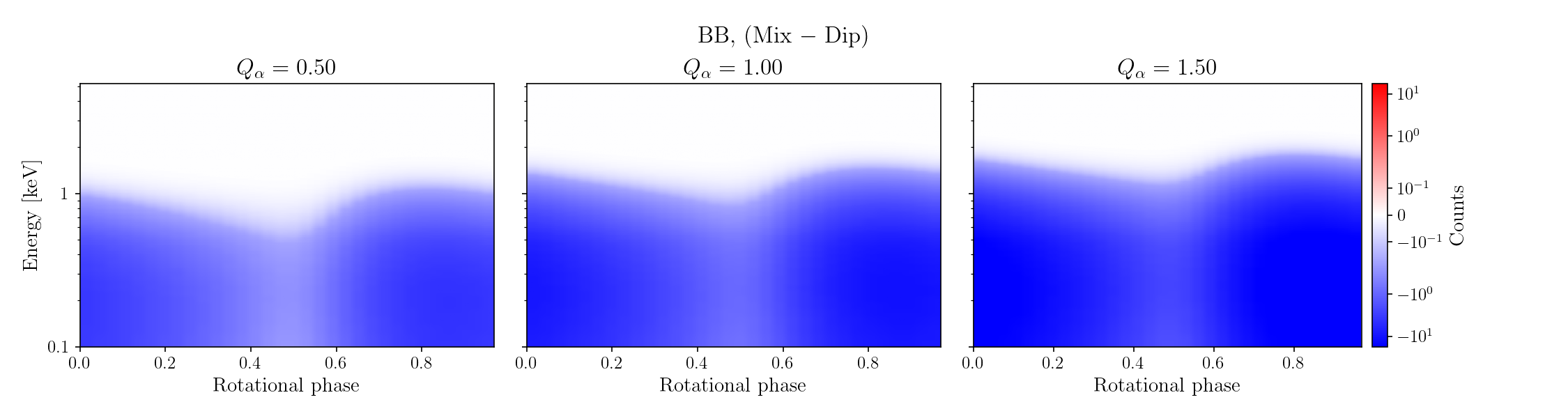}
\caption{Energy-resolved pulse profiles for blackbody (BB) emission, shown as phase--energy maps for the same $Q_\alpha$ sequence
(columns: $Q_\alpha=0.50,1.00,1.50$).
Top row: dipole-only $\Lambda$.
Middle row: \dqmix $\Lambda$.
Bottom row: residual (mix$-$dip), with a zero-centered symmetric-logarithmic color scale.
Horizontal axes show rotational phase, vertical axes show photon energy (keV). The color scale shows the photon counts on each energy-phase bin.}
\label{fig:erpp-bb}
\end{figure*}

\begin{figure*}[t]
\centering
\includegraphics[width=\textwidth]{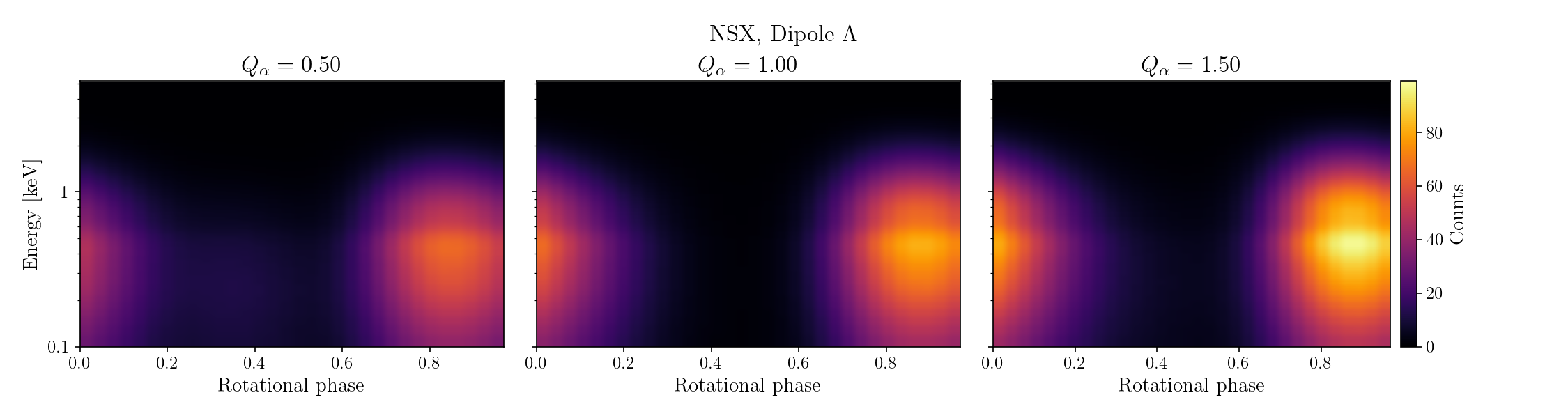}\\
\includegraphics[width=\textwidth]{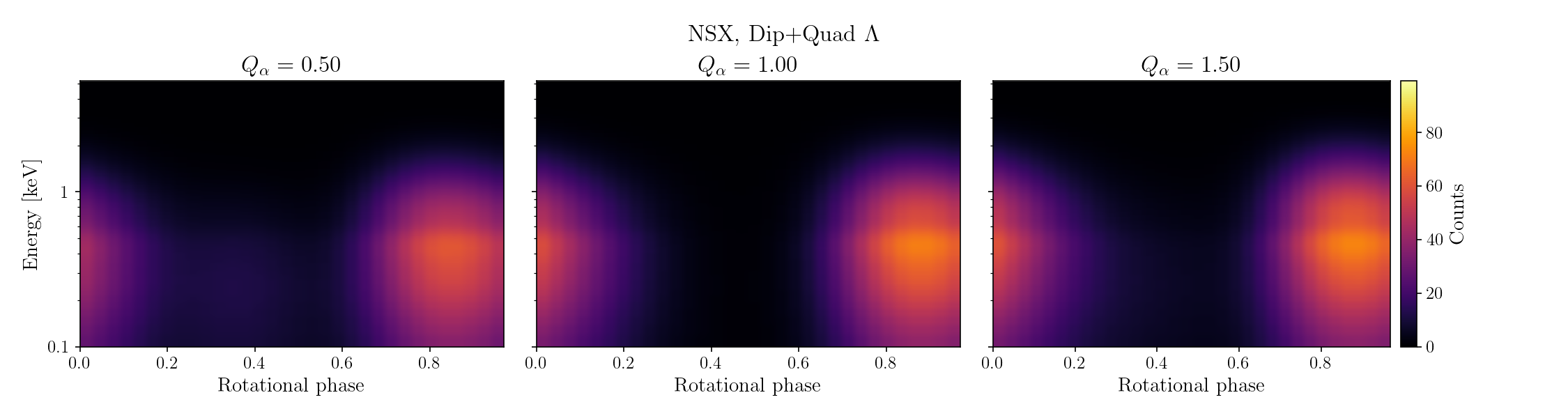}\\
\includegraphics[width=\textwidth]{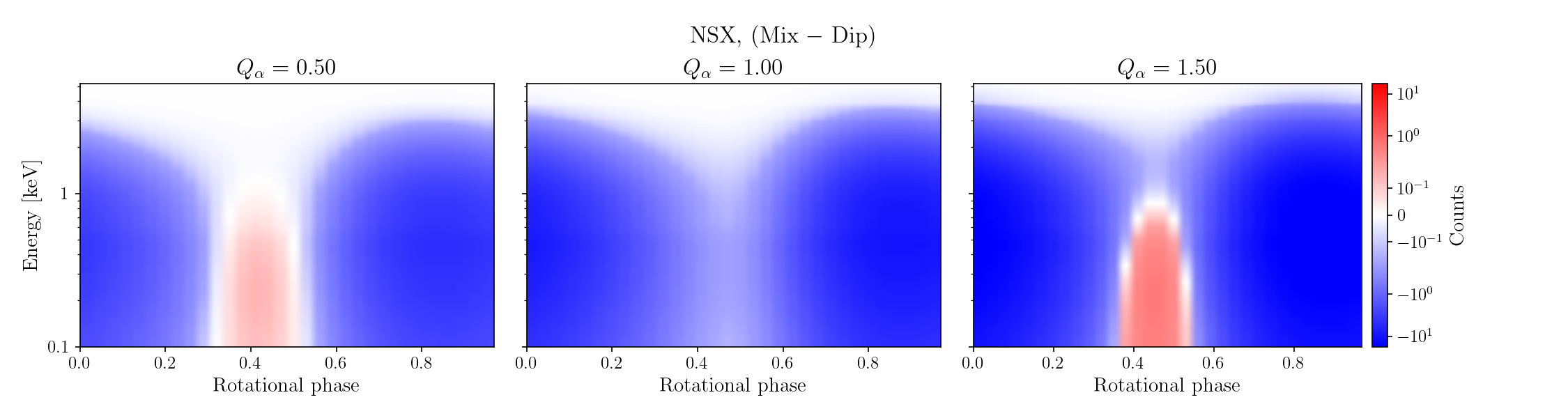}
\caption{Same as Fig.~\ref{fig:erpp-bb}, but for an NSX atmosphere beaming prescription.
Top row: dipole-only $\Lambda$.
Middle row: \dqmix $\Lambda$.
Bottom row: residual (mix$-$dip) with a zero-centered symmetric-logarithmic color scale.}
\label{fig:erpp-nsx}
\end{figure*}

\section{Results}
\label{result}

In this section we quantify how a quadrupolar component modifies the return-current heating pattern and the resulting X-ray observables within the prescription described in Section~\ref{sec:Lambda-derivation}. Because no closed-form Euler potentials are currently available for a fully general (non-axisymmetric) vacuum quadrupole, we restrict the numerical results to the axisymmetric $m=0$ quadrupole superposed with a canonical dipole. In this \dqmix configuration, the Euler potentials and the cap current invariant $\Lambda(\alpha,\beta)$ can be constructed self-consistently from first principles, enabling a controlled comparison between (i) a dipole-only far-zone prescription for $\Lambda$ and (ii) the quadrupole-aware prescription appropriate to the same near-zone field geometry. The quadrupolar contribution modifies the current density on open field lines and therefore changes both the geometry of the heated region and the local heating rate on the polar cap region, producing a different temperature distribution over the emitting patches. Since the observed signal is an integral over these patches with relativistic ray tracing and emission beaming, changes in the surface temperature map propagate directly into the synthetic X-ray light curves and into the two-dimensional, energy-resolved pulse profiles.

Throughout this section we adopt a forward-modeling setup that matches standard choices in contemporary NICER pulse-profile pipelines, so that the results can be used as a direct bias diagnostic for multipolar effects on the computation of X-ray pulse profile. In addition to baseline blackbody (BB) calculations, we repeat the key comparisons with a realistic hydrogen-atmosphere beaming prescription using the NSX model \citep{Ho_2001,Ho_2003}. The central point is that realistic beaming can amplify the observable consequences of modest changes in the hotspot temperature distribution: differences between dipole-only and quadrupole-aware prescriptions for $\Lambda$ that appear moderate at the level of surface maps can become significantly larger in the NICER band, and hence can plausibly bias inferences of neutron-star parameters if a strictly dipolar far-zone normalization is assumed.

\subsection{Mixed dipole–quadrupole hotspots}
\label{subsec:mixed-hotspot}

We first examine how the current-density and temperature distributions on the surface are modified by the quadrupole-aware prescription. The relevant surface quantities are determined by equations~\eqref{eq:Current} and~\eqref{eq:jt}, which together fix the four-current in the near zone once the Euler potentials and $\Lambda(\alpha,\beta)$ are specified. We present current-density maps for three representative values of the quadrupole amplitude parameter $Q_\alpha$, with the corresponding labels shown in the figures. For each $Q_\alpha$ we compare the maps obtained using the dipole-only prescription for $\Lambda$ to those obtained using the quadrupole-aware (\dqmix) prescription derived in Section~\ref{sec:Lambda-DplusQ}, and we then show their relative differences.

To facilitate physical interpretation, we also quote the corresponding far-zone dipole--quadrupole mixing parameter
$\eta$, computed from $Q_\alpha$ via equation~\eqref{eq:eta-Qalpha-lockhart}.
The linear mapping $\eta(Q_\alpha)$ and its dependence on $\nu$ and $R$ are shown in Figure~\ref{fig:qlambda}.
For the implemented value here in this work $(\nu,R)=(400\,{\rm Hz},12\,{\rm km})$ see Table \ref{tab:pp-setup},
$Q_\alpha=0.50,1.00,1.50$ corresponds to $\eta\simeq 0.041,\,0.082,\,0.123$.

Figure~\ref{fig:jsq-maps} displays the current-density distributions for
$Q_\alpha = 0.50,\,1.00,\,1.50$ under the two prescriptions for $\Lambda$.
Comparing the upper and middle rows, we find that both prescriptions produce the same qualitative evolution as $Q_\alpha$ increases: the southern polar hotspot transitions from a relatively faint, nearly circular cap to a broader, band-like (ring-like) structure, while the northern polar cap shrinks in area. This is consistent with the qualitative trends reported by \citet{Gralla2017,Lockhart2019}: increasing the quadrupole moment reorganizes the open-field footprint and the associated return-current distribution. The quantitative differences, however, are substantial. The bottom row of Figure~\ref{fig:jsq-maps} shows the relative difference between the \dqmix and dipole-only prescriptions for $\Lambda$. For each $Q_\alpha$, the current density is modified across essentially the entire open-field cap, and the pattern of enhancement and suppression is not a simple rescaling of the baseline map. In particular, the separations between regions of opposite-sign current are not dramatically shifted, so the gross hotspot extent remains comparable, but the current distribution within the cap is reorganized in a manner that necessarily propagates into the temperature profile. This behavior matches the underlying expectation from the matched-asymptotic construction: the quadrupole contribution decays in the far zone and therefore does not typically change the topology of the open-field region, but it can induce order-unity quantitative changes in the cap-interior current distribution. Neglecting the quadrupolar contribution in $\Lambda$ thus leads to a coherent, spatially structured error in the predicted heating pattern.

Using the prescription described in Section~\ref{subsec:temp-ppm}, we convert the current-density maps into effective temperature distributions on the surface hotspots. Figure~\ref{fig:temp-maps} shows the resulting temperature maps for the same set of quadrupole amplitudes. For the weakest quadrupole considered ($Q_\alpha=0.50$), the overall morphology remains close to the canonical dipolar configuration. As $Q_\alpha$ increases, the southern hotspot becomes progressively broader and more irregular; by $Q_\alpha=1.50$ it exhibits a pronounced band-like structure. These qualitative trends appear in both prescriptions. The difference maps reveal a more nuanced redistribution: for $Q_\alpha=0.50$, the principal temperature enhancement is concentrated on the southern hotspot while the northern cap is uniformly cooler relative to the dipole-only case; for $Q_\alpha=1.00$, the enhancement becomes concentrated in the crescent-like region associated with the strongest return current; and for $Q_\alpha=1.50$, essentially the entire southern cap exhibits a strong enhancement while the northern cap continues to show a net decrease. All temperature maps in Figure~\ref{fig:temp-maps} are normalized to their own maximum, so these patterns reflect genuine redistributions of heating rather than a trivial overall rescaling.

The immediate implication is that the predicted light curves and energy-resolved pulse profiles should differ materially between the two prescriptions: the quadrupole-aware $\Lambda$ changes both the shape of the emitting regions and the relative weighting of hot and cool patches as a function of rotational phase. In the next subsection we quantify this impact at the level of band-integrated light curves and two-dimensional, energy-resolved pulse profiles, and we show that omitting the quadrupolar contribution to the flux--current functional $\Lambda$ can produce large, phase-dependent discrepancies once realistic beaming is included.

\subsection{Quantifying the quadrupole contribution to light curves and pulse profiles}
\label{subsec:lc-erpp}

The X-ray light curve and energy-resolved pulse profile are the primary observables used by NICER for neutron-star parameter inference. It is therefore essential to quantify how the quadrupole-aware prescription modifies these observables relative to a dipole-only prescription for $\Lambda$, holding fixed the same underlying near-zone field geometry.

For all light-curve and energy-resolved pulse-profile calculations in this subsection, we adopt a single fiducial neutron-star and observer configuration summarized in Table~\ref{tab:pp-setup}. We fix $M=1.4\,M_\odot$, $R=12\,$km, and $\nu=400\,$Hz, set the observer inclination to $i=1.2\,\mathrm{rad}$, and adopt a source distance $d=0.3\,\mathrm{kpc}$. The simulated exposure time is $T_{\rm exp}=5\times10^5\,\mathrm{s}$, the surface temperature map is discretized with ${\tt Nside}=64$ (HEALPix), and the rotational phase integration uses ${\tt fine\_phase}=32\times 6$ sub-bins and ${\tt points\_per\_bin}=2$ integration points per observed phase bin, corresponding to the standard numerical configuration \texttt{setting = "std"} in our implementation. As demonstrated in \citet{GPU_ppm}, this configuration is sufficient to resolve complex hotspot structure including sharp edges, while remaining numerically stable. Unless otherwise stated, each panel uses either blackbody (BB) emission or NSX atmosphere emission, as indicated in the figure labels.

We begin with the band-integrated light curve computed over $0.1$--$5.2125$~keV, matching the energy range adopted in our forward-model configuration for NICER-like comparisons. Figure~\ref{fig:lc-panels} shows the resulting light curves for several values of $Q_\alpha$. In each panel, the flux is normalized to the maximum among the curves shown in that panel, facilitating a direct comparison of relative amplitudes and phase structure. The lower sub-panels show the relative flux difference between the dipole-only and quadrupole-aware prescriptions for $\Lambda$, evaluated separately for BB and NSX.

A clear trend emerges: for all $Q_\alpha$, switching from BB to NSX amplifies the discrepancy between dipole-only and quadrupole-aware predictions. The discrepancy also grows systematically with increasing $Q_\alpha$. For $Q_\alpha=0.50$ the differences are at the few-percent level, while by $Q_\alpha=1.50$ the BB light curves exhibit peak differences at the $\sim 10\%$ level and the NSX light curves reach peak differences of order $\sim 30\%$. The discrepancies are concentrated primarily near the flux maxima rather than the troughs, which makes them particularly relevant for inference because pulse-profile fits are often most sensitive to the detailed shape and height of the peaks. Crucially, realistic atmosphere beaming does not wash out the quadrupole-induced differences. Instead, it renders them more pronounced and more structured in phase. This demonstrates that neglecting the quadrupolar contribution in the far-zone mixing encoded by $\Lambda$ can lead to materially biased light-curve predictions even when the qualitative hotspot morphology appears similar.

We next examine the two-dimensional, energy-resolved pulse profiles. Figure~\ref{fig:erpp-bb} shows the results for BB emission. Each panel compares the dipole-only and \dqmix prescriptions and displays their photon-count difference as a function of rotational phase and photon energy. As $Q_\alpha$ increases from $0.50$ to $1.00$ and $1.50$, the overall morphology of the pulse profiles, including the number of peaks and their phase locations are largely preserved. However, the peak brightness changes systematically. In these examples, the dipole-only prescription predicts higher photon counts than the quadrupole-aware prescription, consistent with the trend observed in the band-integrated light curves. The excess is concentrated near pulse maxima, phases and energies that are bright in the baseline model are preferentially overestimated when the quadrupole is neglected. Quantitatively, the relative count excess extends over most of the rotational phase and across the energy channels that carry substantial signal. The magnitude of the discrepancy increases with $Q_\alpha$, becoming severe for $Q_\alpha=1.50$, as expected when the quadrupole carries a larger fraction of the far-zone field. Thus, even for quadrupole amplitudes that remain compatible with a broadly dipole-like far-zone morphology, omitting the quadrupole contribution in $\Lambda$ produces a systematic, phase-dependent bias in the predicted energy-resolved signal.

Figure~\ref{fig:erpp-nsx} repeats the comparison using NSX atmosphere emission. Atmosphere beaming reshapes the energy dependence of the pulse profile, shifting a larger fraction of the modulated signal to higher energies relative to BB emission. More importantly for our purpose, the qualitative conclusion remains unchanged: including the quadrupolar contribution in $\Lambda$ produces a clear, phase-dependent modification of the predicted counts, and the magnitude of this modification grows with $Q_\alpha$. In the NSX case, the dominant effect remains a reduction of photon counts near the main peak(s) when the quadrupole-aware prescription is used, mirroring the bolometric trends. In addition, for some $Q_\alpha$ values (notably $Q_\alpha=0.50$ and $1.50$ in our examples), the interpulse region can show a modest increase in photon counts in the quadrupole-aware model. This behavior is consistent with the temperature-map differences: the quadrupole redistributes heating within the cap, and atmosphere beaming can enhance the visibility of particular surface patches in specific energy bands and phase intervals. The interpulse enhancements occur primarily where the baseline photon counts are low, so the overall effect remains a net suppression near the principal maxima, but the phase-dependent structure introduced by the atmosphere underscores that realistic beaming can convert relatively subtle hotspot redistributions into distinctly nontrivial systematic bias.

The opposite response of the two polar caps can be understood directly from the structure of the \dqmix field geometry. Our starting point is the combined Euler potential, equation~\eqref{eq:alpha-mu-Q}, in which the quadrupolar contribution carries an additional factor of $\cos\theta'$. This factor is odd under reflection across the magnetic equator, $\theta'\rightarrow\pi-\theta'$, whereas the dipolar term $\propto \sin^2\theta'$ is even. For a fixed sign of $Q_\alpha$, the quadrupole therefore enters with opposite sign in the two hemispheres, on one polar cap it reinforces the dipolar contribution to the open-flux structure, while on the other it partially cancels it.

Through the linear superposition embodied by the flux--current functional,
\begin{equation}
\Lambda(\chi,\beta)
=
(2\Omega)\big[\Lambda_D(\chi,\beta)+\eta\,\Lambda_Q(\chi,\beta)\big],
\label{eq:Lambda-mixed-explain}
\end{equation}
and the return-current expressions in equation~\eqref{eq:Current}, this hemispheric sign flip translates into opposite trends in the local current density and hence in the heating rate. In the sign convention adopted in our numerical examples, $Q_\alpha>0$ is chosen such that the quadrupolar contribution is aligned with the dipole on the southern cap and anti-aligned on the northern cap. The southern cap therefore exhibits a net reinforcement of the return current over a finite range of colatitudes, producing the enhanced, ring-like temperature structures seen in Figure~\ref{fig:temp-maps}. On the northern polar cap, the quadrupole contribution tends to weaken the dipolar return current over most of the open region, yielding an overall reduction in the temperature map.

Because $\Lambda_Q(\chi,\beta)$ contains higher azimuthal structure (through the $J_1$ and $J_2$ modes and their angular dependence), the reinforcement and cancellation are not spatially uniform: different regions of the cap can be enhanced or depleted, which explains the structured difference maps in both current density and temperature. The north--south asymmetry in the thermal response is therefore a direct consequence of the odd parity of the quadrupolar contribution in the mixed Euler potential and of the linear way in which dipolar and quadrupolar components enter the cap current invariant and the return-current prescription.

\section{Discussion}
\label{sec:discussion}

Several simplifying assumptions and limitations of the present work should be emphasized. First, although we derived an analytic expression for $\Lambda$ for an arbitrary quadrupole tensor $Q_{ij}$, the numerical hotspot and light-curve demonstrations in this paper were restricted to a dipole plus axisymmetric ($m=0$) quadrupole. This restriction reflects the current absence of a convenient closed-form representation of the Euler potentials $(\alpha,\beta)$ for a generic non-axisymmetric quadrupole. Implementing fully general quadrupolar geometries in pulse-profile calculations will likely require numerically tabulated Euler potentials or asymptotic representations (e.g., controlled analytic fits or machine-learned emulators). The general expression $\Lambda(Q^{(\Omega)}_{2m})$ derived here is formulated to be immediately usable once such representations of $(\alpha,\beta)$ become available, but we have not yet deployed it in end-to-end pulse-profile computations for fully non-axisymmetric quadrupoles. This is deferred for future work.

Second, our construction is anchored in the matched-asymptotic expansion in the small parameter $\epsilon = R_\star/R_{\rm LC} = \Omega R_\star/c$, and it retains only the leading-order cap physics in this ordering. In this regime the open-field region occupies a small patch on the stellar surface that can be conformally mapped to a unit disk, and the cap--interior current is treated as a harmonic function supplemented by a thin, unresolved return-current layer along the separatrix. In the mixed dipole--quadrupole examples considered here, the open cap remains small in angular extent, even though the \emph{heating pattern} within that cap can become strongly concentrated in an annular (ring-like) band near the rim. The small-cap formulation therefore remains a useful organizing principle for the scenarios explored in this work, but it does not capture effects that arise when the open-field region becomes strongly distorted, develops complex topology, or approaches order-unity angular scales. Moreover, we neglect higher-order corrections in $\epsilon$, departures from the idealized conformal-cap description, and finite-thickness structure of the separatrix layer. Quantifying the impact of these effects on high-precision pulse-profile modeling remains an important direction for future work.

Third, we model the near-star magnetosphere as perfectly force-free and focus on the large-scale mapping between far-zone current structure and polar-cap return currents. Microphysical ingredients like pair creation, resistivity, gap formation, surface conductivity, and possible deviations from ideal force-free conditions, enter only through a deliberately simple heating prescription that maps the local current to an effective temperature. In particular, we do not attempt to resolve particle-acceleration regions, non-thermal emission components, or the feedback of pair cascades on the global current closure. While this level of description is appropriate for isolating the \emph{geometric} imprint of multipolar fields on thermal hotspots, a more realistic microphysical treatment could modify the quantitative relation between $J_\parallel$  and the emergent surface flux, and thereby change the detailed amplitude of the quadrupole-induced signatures.

Fourth, our forward-modeling tests were carried out for a single fiducial neutron-star mass, radius, spin, and viewing geometry, and for a limited sequence of quadrupole amplitudes. We have not explored the broader parameter space of compactness, inclination, obliquity, nor have we folded the synthetic signals through a full observational pipeline. Interstellar absorption, instrumental response, background, and other data characteristics were simplified or omitted to isolate the theoretical effect of magnetic geometry. Consequently, while the relative differences reported here between dipole-only and quadrupole-aware prescriptions provide a robust indicator of potential modeling bias, translating these differences into quantitative biases on inferred mass, radius, and hotspot configuration parameters will require a full Bayesian analysis with realistic data modeling.

Finally, the harmonic-map viewpoint and the representation-theory logic used here strongly suggest that analogous constructions can be pursued for higher $\ell$, but we have not carried out those extensions in this work. The present paper should therefore be viewed as a controlled, physically transparent demonstration that even the lowest non-dipolar multipole can already produce appreciable changes in hotspot structure and X-ray observables, at a level relevant for precision pulse-profile modeling.

\section{Conclusion and Outlook}
\label{conclusion}
We have derived analytic expressions for the conserved field-aligned current invariant $\Lambda(\alpha,\beta)$ in oblique, force-free pulsar magnetospheres with mixed dipole--quadrupole geometries. This work establishes a consistent analytic connection between magnetospheric multipoles and magnetospheric current density starting from the stationary, ideal force-free fields formulation, we showed that the cap-interior current invariant behaves harmonically to leading order on the open polar cap. This property allows us to construct an analytic Bessel-harmonic solution that naturally preserves the correct physical behavior near the magnetic pole. This construction recovers the familiar inclined-dipole form of \citet{Gralla2017} as the minimal harmonic solution matched to the far-zone current, providing a first-principles rationale for the semi-analytic expression used in earlier work. We then extended the derivation to an axisymmetric ($m=0$) quadrupole and, more generally, to an arbitrary quadrupolar tensor $Q_{ij}$, obtaining a compact expression for $\Lambda$ in terms of the spin-frame components $Q^{(\Omega)}_{2m}$ and the Euler potentials. Within the class of minimal, regular, linear, rotationally covariant constructions, $\Lambda$ is uniquely fixed up to a single overall normalization determined by near--far matching. These results yield explicit formulae for the cap current distribution and the associated surface temperature map in mixed dipole--quadrupole geometries, without requiring a global force-free simulation for each choice of parameters.

To evaluate the physical implications of this framework on surface heating, we specialized to a mixed dipole--quadrupole configuration, for which the Euler potentials are available in closed form, and we explored a sequence of increasing quadrupole amplitudes $Q_\alpha$. In this mixed geometry, the quadrupolar contribution reorganizes the return-current pattern across the caps. The southern cap develops strongly nonuniform heating that can become concentrated into a band-like structure near the rim, while the northern cap typically shrinks and exhibits an overall reduction in heating. The resulting temperature maps inherit these redistributions, producing structured departures from the dipole-only baseline even when the gross hotspot configuration remains similar. These qualitative trends are consistent with earlier force-free studies, but here they follow from analytic expressions that can be evaluated at negligible computational cost.

We then propagated these temperature maps through a relativistic ray-tracing pipeline, adopting a fiducial millisecond pulsar ($M=1.4\,M_\odot$, $R=12\,\mathrm{km}$, $\nu=400\,\mathrm{Hz}$) with fixed viewing geometry. The computation accounted for general-relativistic light bending, gravitational redshift, and frame dragging, as well as special-relativistic Doppler boosting, aberration, and spin-induced stellar oblateness. We computed NICER-band light curves and energy-resolved pulse profiles assuming either isotropic blackbody emission or a realistic hydrogen-atmosphere beaming model. Across the explored parameter sequence, the predictions of the quadrupole-aware framework depart systematically from those obtained using the dipole-only prescription for $\Lambda$. While isotropic blackbody light curves show peak deviations of $\sim 10\%$ by $Q_\alpha\simeq 1.5$, the inclusion of realistic atmosphere beaming amplifies these phase-localized discrepancies to $\sim 30\%$ near flux maxima. This demonstrates that if an MSP magnetosphere contains even a modest quadrupolar component in the matching region, relying on a dipole-only far-zone normalization leads to systematic biases in the predicted X-ray signal that far exceed the precision requirements of modern pulse-profile modeling.

The analytic framework developed here provides a direct connection between multipolar magnetic geometry and physics-motivated pulse-profile modeling, and it suggests several concrete directions for future work. On the theory side, the most immediate step is to combine the general quadrupole expression $\Lambda(Q^{(\Omega)}_{2m})$ with a usable representation of the Euler potentials for non-axisymmetric quadrupoles, whether via numerical tabulation or machine-learning-based asymptotic models. Once $(\alpha,\beta)$ are available in such a form, the formalism here can be used without modification to compute polar-cap currents and hotspot maps for generic quadrupole orientations and phases, providing a template for extensions to higher multipoles.

On the pulse profile modeling side, a natural next step is to integrate the quadrupole-aware $\Lambda(\alpha,\beta)$ prescription directly into the GPU-accelerated pulse-profile modeling framework of \citet{GPU_ppm}. Because $\Lambda$ is analytic and inexpensive to evaluate, it is well suited for large-scale parameter exploration and Bayesian inference with physics-motivated hotspot models parameterized by underlying dipole--quadrupole degrees of freedom. In combination with the strategies explored in \citet{Huang:2025_hotspot}, this would enable controlled quantification of the bias incurred when a truly multipolar MSP is analyzed under idealized dipolar assumptions, and it would allow direct constraints to be placed on the quadrupolar content of individual sources.

More broadly, this work establishes that incorporating quadrupole-aware polar-cap physics is a fundamental requirement for physical consistency, rather than merely a refinement for high-precision regimes. By demonstrating that the dipole-only far-zone assumption introduces systematic physical biases that cannot be ignored, we show that implementing this self-consistent framework is essential for robust constraints in any physics-motivated pulse-profile modeling effort. By replacing \emph{ad hoc} hotspot parameterizations with analytic connections between the far-zone magnetic structure, polar-cap currents, and surface heating, the framework developed here offers a rigorous pathway to bringing magnetic-geometry systematics under theoretical control. Looking forward, as the community moves toward combining X-ray pulse profiles with radio and $\gamma$-ray constraints, the analytic $\Lambda$ solutions derived here provide a unified physical foundation for the multiwavelength modeling of multipolar neutron-star magnetospheres.
\begin{acknowledgments}
The author acknowledges support from NSF Grant AST-2308111. The author thanks Roger Romani and Alexander Chen for insightful discussions regarding gamma-ray constraints on possible quadrupole contributions to the magnetosphere.
\end{acknowledgments}

\section*{Appendix A}
\label{appendixA: rms_derivation}
Here we provide a short derivation of the relation between the near-zone quadrupole coefficient $Q_\alpha$ and the far-zone quadrupole-to-dipole field ratio $\eta$, and we show explicitly where equation~\eqref{eq:eta-Qalpha-lockhart} comes from.

Starting from the mixed Euler potential used in the main text, equation~\eqref{eq:alpha-mu-Q}, the leading-order vacuum limit at $r\gg R_\star$ can be written, up to overall GR radial factors, as
\begin{equation}
\alpha(r,\theta)
\simeq
\mu\,\hat{R}_{1}^{>}(r)\,\sin^2\theta
+
Q_\alpha\,\hat{R}_{2}^{>}(r)\,\sin^2\theta\,\cos\theta,
\end{equation}
where $\hat{R}_{\ell}^{>}(r)\equiv R_{\ell}^{>}(r)/R_{\ell}^{>}(R_\star)$ are the normalized exterior radial eigenfunctions.  In the far zone, these satisfy the usual vacuum scalings $\hat{R}_{1}^{>}(r)\propto R_\star/r$ and $\hat{R}_{2}^{>}(r)\propto (R_\star/r)^2$, so we may write, to leading order in $R_\star/r$,
\begin{equation}
\alpha(r,\theta)
\simeq
\mu\,\frac{R_\star}{r}\,\sin^2\theta
+
Q_\alpha\,\Big(\frac{R_\star}{r}\Big)^{\!2}\sin^2\theta\,\cos\theta.
\label{eq:alpha-appendix}
\end{equation}
For an axisymmetric poloidal field, the Euler potential determines the magnetic field via
\begin{equation}
B_r
=
\frac{1}{r^2\sin\theta}\frac{\partial\alpha}{\partial\theta},
\qquad
B_\theta
=
-\frac{1}{r\sin\theta}\frac{\partial\alpha}{\partial r},
\label{eq:Br-Bt-appendix}
\end{equation}
which is equivalent to using $\mathbf{B}=\nabla\alpha\times\nabla\beta$ with $\beta=\phi$.

It is convenient to rewrite equation~\eqref{eq:alpha-appendix} as
\begin{equation}
\begin{aligned}
\alpha(r,\theta)\simeq\;&
\frac{\tilde{\mu}\,\sin^2\theta}{r}
+
\frac{\tilde{Q}\,\sin^2\theta\,\cos\theta}{r^2},
\\[6pt]
&\tilde{\mu}\equiv \mu R_\star,
\qquad
\tilde{Q}\equiv Q_\alpha R_\star^2.
\end{aligned}
\end{equation}

so that the calculation reduces to the standard vacuum multipole algebra.

For the dipole part $\alpha_D=\tilde{\mu}\sin^2\theta/r$ one finds
\begin{align}
\frac{\partial\alpha_D}{\partial\theta}
&=
\frac{2\tilde{\mu}\sin\theta\cos\theta}{r}, &
\frac{\partial\alpha_D}{\partial r}
&=
-\frac{\tilde{\mu}\sin^2\theta}{r^2},
\end{align}
and hence, from equation~\eqref{eq:Br-Bt-appendix},
\begin{equation}
B_{r,D}
=
\frac{2\tilde{\mu}\cos\theta}{r^3},
\qquad
B_{\theta,D}
=
\frac{\tilde{\mu}\sin\theta}{r^3}.
\end{equation}
Thus
\begin{equation}
B_D^2 \equiv B_{r,D}^2 + B_{\theta,D}^2
=
\frac{\tilde{\mu}^2}{r^6}\bigl(4\cos^2\theta + \sin^2\theta\bigr).
\end{equation}
The rms dipolar field at radius $r$ is defined by
\begin{equation}
\big\langle B_D^2 \big\rangle(r)
\equiv
\frac{1}{4\pi}\int B_D^2\,\mathrm{d}\Omega,
\end{equation}
which evaluates to
\begin{equation}
\big\langle B_D^2 \big\rangle(r)
=
\frac{2\tilde{\mu}^2}{r^6}.
\label{eq:BD-rms-appendix}
\end{equation}

For the quadrupole part $\alpha_Q = \tilde{Q}\sin^2\theta\,\cos\theta/r^2$ one obtains
\begin{equation}
\begin{aligned}
\frac{\partial\alpha_Q}{\partial\theta}
&=
\frac{\tilde{Q}}{r^2}
\bigl(
2\sin\theta\cos^2\theta
- \sin^3\theta
\bigr),
\\[6pt]
\frac{\partial\alpha_Q}{\partial r}
&=
-\frac{2\tilde{Q}\sin^2\theta\cos\theta}{r^3}.
\end{aligned}
\end{equation}

and therefore
\begin{equation}
B_{r,Q}
=
\frac{\tilde{Q}}{r^4}\bigl(2-3\sin^2\theta\bigr),
\qquad
B_{\theta,Q}
=
\frac{\tilde{Q}}{r^4}\sin(2\theta).
\end{equation}
This implies
\begin{align}
B_Q^2
&\equiv
B_{r,Q}^2+B_{\theta,Q}^2
=
\frac{\tilde{Q}^2}{r^8}
\Big[
\bigl(2-3\sin^2\theta\bigr)^2 + \sin^2(2\theta)
\Big]
\nonumber\\
&=
\frac{\tilde{Q}^2}{r^8}
\bigl(
5\sin^4\theta - 8\sin^2\theta + 4
\bigr),
\end{align}
where we used $\sin(2\theta)=2\sin\theta\cos\theta$.  The rms quadrupolar field is then
\begin{equation}
\begin{aligned}
\big\langle B_Q^2 \big\rangle(r)
&=
\frac{1}{4\pi}\int B_Q^2\,\mathrm{d}\Omega\\
&=
\frac{\tilde{Q}^2}{2\,r^8}
\int_0^\pi \!\mathrm{d}\theta\,\sin\theta\,
\bigl(5\sin^4\theta - 8\sin^2\theta + 4\bigr)
\\[6pt]
&=
\frac{4}{3}\,\frac{\tilde{Q}^2}{r^8}.
\end{aligned}
\label{eq:BQ-rms-appendix}
\end{equation}

The rms quadrupole-to-dipole field ratio at radius $r$ is defined by
\begin{equation}
\left(\frac{B_Q}{B_D}\right)_{\!\rm RMS}(r)
\equiv
\sqrt{
\frac{\langle B_Q^2\rangle(r)}{\langle B_D^2\rangle(r)}
}.
\end{equation}
Using equations~\eqref{eq:BD-rms-appendix} and~\eqref{eq:BQ-rms-appendix}, we obtain
\begin{align}
\left(\frac{B_Q}{B_D}\right)_{\!\rm RMS}(r)
&=
\sqrt{
\frac{\frac{4}{3}\tilde{Q}^2/r^8}{2\tilde{\mu}^2/r^6}
}
=
\sqrt{\frac{2}{3}}\,
\frac{\tilde{Q}}{\tilde{\mu}}\,
\frac{1}{r}
=
\frac{2}{\sqrt{6}}\,
\frac{Q_\alpha}{\mu}\,
\frac{R_\star}{r}.
\end{align}
Thus the numerical factor $2/\sqrt{6}$ arises directly from the angular averages of $|\mathbf{B}_D|^2$ and $|\mathbf{B}_Q|^2$ over the sphere.

Finally, evaluating this ratio at the light cylinder $r=R_{\rm LC}=c/\Omega$ and writing $R_{\rm LC}=R_\star/\epsilon$ with $\epsilon\equiv \Omega R_\star/c$, we obtain
\begin{equation}
\eta
\;\equiv\;
\frac{B_Q}{B_D}\bigg|_{R_{\rm LC}}
\;\approx\;
\left(\frac{B_Q}{B_D}\right)_{\!\rm RMS}(R_{\rm LC})
=
\frac{2}{\sqrt{6}}\,
\frac{Q_\alpha}{\mu}\,\epsilon,
\end{equation}
which is the expression quoted in equation~\eqref{eq:eta-Qalpha-lockhart}.  Any residual order-unity differences arising from the full GR radial eigenfunctions or from alternative angular weightings can be absorbed into the geometric factor $\kappa$ introduced in equation~\eqref{eq:A2-cal}.

\section*{Appendix B}\label{app:derive_2d}

{
We take as starting point the standard leading-order force-free result that the current function is constant along
magnetic field lines, so that $\Lambda=\Lambda(\alpha,\beta)$ in terms of Euler potentials $(\alpha,\beta)$ labeling
open field lines. This field-line constancy is a three-dimensional, first-order constraint \emph{along}
field lines; by itself it does not determine how $\Lambda$ varies \emph{between} neighboring open field lines across the
polar cap.} {The matched-asymptotic assumptions used in the main text motivate posing a smooth interior boundary-value problem
on the cap, but they do not uniquely fix $\Lambda(\alpha,\beta)$.
Accordingly, equation~\eqref{eq:conservation_W} is introduced as an explicit additional interior closure ansatz,
whose mathematical form and motivation are given below.
}

{
On the stellar surface, let $x^a$ $(a=1,2)$ be coordinates with induced metric $g_{ab}$, determinant $g$, and associated
covariant derivative $\nabla_\perp$. The normal magnetic field can be written in terms of Euler potentials as
}
\begin{equation}
B_n \;=\;\frac{1}{\sqrt{g}}\,\epsilon^{ab}\,\partial_a\alpha\,\partial_b\beta\,,
\label{eq:BnEuler_app}
\end{equation}
{
where $\epsilon^{ab}$ is the Levi-Civita tensor density on the surface. Consequently, the natural ``flux-tube counting''
measure on the polar cap is
}
\begin{equation}
d\mu \;\equiv\; B_n\,\sqrt{g}\,d^2x\,,
\label{eq:FluxTubeMeasure_app}
\end{equation}
{
i.e.\ surface area weighted by the local density of open flux tubes. We therefore introduce a strictly positive scalar
weight $\mathcal{W}$ proportional to this open-flux density,
}
\begin{equation}
\mathcal{W}\;\propto\;B_n\,,
\label{eq:Wdef_app}
\end{equation}
{
with an overall normalization irrelevant for our later temperature normalization. Note that $\mathcal{W}$ is a scalar
defined on the \emph{two-dimensional} stellar surface; even if $\mathcal{W}\propto B_n$, it does \emph{not} convert the
intrinsic surface derivative $\nabla_\perp$ into the field-line derivative $\mathbf{B}\!\cdot\!\nabla$.
}

{
To close the interior problem we adopt the matched-asymptotic current-closure picture emphasized by \citet{Gralla2017}
and \citet{Lockhart2019}: (i) the open-cap interior is smooth and contains no resolved current sources or sinks at the
order retained, while (ii) global current closure is achieved in a thin, unresolved return-current layer localized on
the separatrix, which supplies boundary data for the smooth interior continuation. Under these assumptions,
$\Lambda(\alpha,\beta)$ is determined in the cap interior by an intrinsic two-dimensional constraint together with
separatrix boundary data.
}

{
In the smooth interior, ``no resolved sources/sinks'' is naturally expressed as a local conservation law on the
two-surface: there exists a surface flux $F^a$ whose divergence vanishes,
$\nabla_a F^a=0$. Locality and rotational invariance on the cap imply that, to leading order in surface gradients,
$F^a$ may depend on $\Lambda$ only through its first derivatives and must take the isotropic form
$F^a=-\kappa(x)\,g^{ab}\nabla_b\Lambda$ with $\kappa(x)>0$. Requiring the resulting operator to be self-adjoint with
respect to the flux-tube measure $d\mu\propto B_n\sqrt{g}\,d^2x$ fixes $\kappa\propto\mathcal{W}\propto B_n$ (up to an
overall normalization). We therefore define the weighted surface flux as
}
\begin{equation}
F^a \;\equiv\; -\,\mathcal{W}\,g^{ab}\,\nabla_b\Lambda\,,
\label{eq:Fdef_app}
\end{equation}
{
so that the interior conservation statement becomes the intrinsic two-dimensional divergence constraint
$\nabla_a F^a=0$, i.e.
}
\begin{equation}
\nabla_\perp\cdot\!\left(\mathcal{W}\,\nabla_\perp\Lambda\right)=0\,.
\label{eq:WeightedLaplace_app}
\end{equation}
{
Equation~\eqref{eq:WeightedLaplace_app} is a second-order elliptic equation intrinsic to the stellar two-surface: it
governs cross-cap structure given separatrix boundary data. It is logically independent of the three-dimensional
field-line constancy condition $\mathbf{B}\!\cdot\!\nabla\Lambda=0$.
}

{
It is sometimes useful to express the same interior closure in a variational form. Among all smooth functions on the
cap that match the separatrix boundary data, the solution of equation~\eqref{eq:WeightedLaplace_app} is equivalently the unique
minimizer of the weighted Dirichlet functional
}
\begin{equation}
\mathcal{I}[\Lambda]
\;\equiv\;
\frac{1}{2}\int_{\rm cap}\mathcal{W}\,g^{ab}\,(\nabla_a\Lambda)(\nabla_b\Lambda)\,\sqrt{g}\,d^2x\,,
\label{eq:DirichletFunctional_app}
\end{equation}
{
which is strictly convex for $\mathcal{W}>0$. Varying \eqref{eq:DirichletFunctional_app} at fixed boundary values
($\delta\Lambda=0$ on the separatrix) gives
}
\begin{align}
\delta\mathcal{I}
&=\int_{\rm cap}\mathcal{W}\,g^{ab}\,(\nabla_a\Lambda)(\nabla_b\delta\Lambda)\,\sqrt{g}\,d^2x \nonumber\\
&=-\int_{\rm cap}\nabla_a\!\left(\mathcal{W}\,g^{ab}\nabla_b\Lambda\right)\delta\Lambda\,\sqrt{g}\,d^2x
\;+\;\text{(boundary term)}\,,
\end{align}
{
and the boundary term vanishes by the Dirichlet condition. Requiring $\delta\mathcal{I}=0$ for arbitrary interior
$\delta\Lambda$ yields equation~\eqref{eq:WeightedLaplace_app}, equivalently
}
\begin{equation}
\frac{1}{\sqrt{g}}\,\partial_a\!\left(\sqrt{g}\,\mathcal{W}\,g^{ab}\,\partial_b\Lambda\right)=0\,.
\end{equation}
{
Finally, equation~\eqref{eq:WeightedLaplace_app} should not be confused with resistive surface-current closure equations (e.g., \citealt{Beskin2010}), such heating models lie outside the scope of the
present force-free/matched-asymptotic closure picture.
}

\bibliography{sample7}{}
\bibliographystyle{aasjournalv7}

\end{document}